# Mixed Quantum/Classical Theory for Rotational Energy Exchange in Symmetric-Top-Rotor + Linear-Rotor Collisions and a Case Study of $ND_3$ + $D_2$ System

Carolin Joy, Bikramaditya Mandal, Dulat Bostan, and Dmitri Babikov*

**Abstract:** The extension of mixed quantum/classical theory (MQCT) to describe collisional energy transfer is developed for symmetric-top-rotor + linear-rotor system type and is applied to $ND_3$ + $D_2$. State-to-state transition cross sections are computed in a broad energy range for all possible processes: when both $ND_3$ and $D_2$ molecules are excited or both are quenched, when one is excited while the other is quenched and vice versa, when $ND_3$ state changes its parity while $D_2$ is excited or quenched, and when $ND_3$ is excited or quenched while $D_2$ remains in the same state, ground or excited. In all these processes the results of MQCT are found to approximately satisfy the principle of microscopic reversibility. For a set of sixteen state-to-state transitions available from literature for collision energy 800 $cm^{-1}$ the values of cross sections predicted by MQCT are within 8% of accurate full-quantum results. A useful time-dependent insight is obtained by monitoring the evolution of state populations along MQCT trajectories. It is shown that, if before the collision, $D_2$ is in its ground state, the excitation of $ND_3$ rotational states proceeds through a two-step mechanism in which the kinetic energy of molecule-molecule collision is first used to excite $D_2$ and only then is transferred to the excited rotational states of $ND_3$. It is found that both potential coupling and Coriolis coupling play important roles in $ND_3$ + $D_2$ collisions.

**TOC Image:** Evolution of state populations in a typical MQCT trajectory for $ND_3$ + $D_2$ collision.

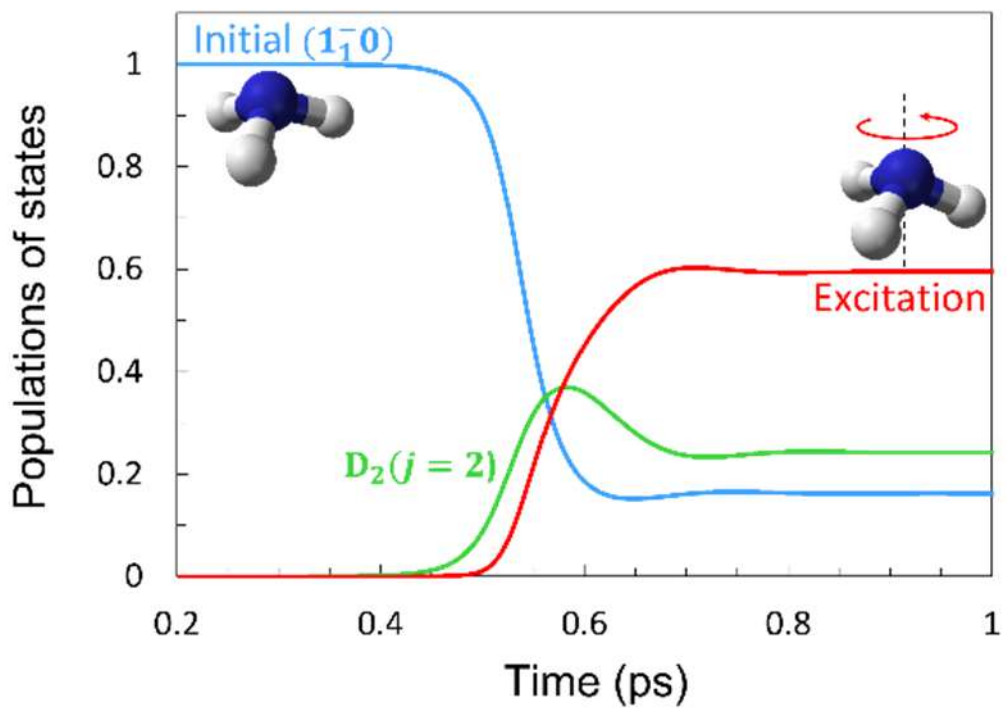

* Author to whom all correspondence should be addressed; electronic mail: dmitri.babikov@mu.edu
*Chemistry Department, Wehr Chemistry Building, Marquette University, Milwaukee, Wisconsin 53201-1881, USA*

## I. INTRODUCTION

Collisional energy transfer (CET) is a key step in many physical and chemical phenomena that involve gas-phase molecules. In this process, energy is exchanged between translational and internal degrees of freedom of the molecules, such as vibrations and/or rotations. CET plays a significant role in probing and discerning molecular interactions. It is crucial in comprehending many processes such as combustion,[1–4] recombination reactions,[5–7] molecule-molecule[8–11] and molecule-surface inelastic scattering,[12–15] astrochemistry,[16,17] atmospheric chemistry,[18–20] photo-chemistry,[21,22] and chemistry at ultracold temperatures.[23–26]

Over the years, many experimental and theoretical studies have paved the way to broaden our understanding of the energy transfer that occur during the collision of two molecules.[27–31] The most widely applied theoretical methods to study CET are the classical trajectory method, often referred to as the quasi-classical trajectory (QCT) method,[32–35] and the full quantum theory of molecular scattering, known as a coupled-channel (CC) formalism [36–41] in which both the collision process and the internal states of molecules are treated using time-independent Schrodinger equation. Despite being physically indispensable and accurate, the numerical effort attributed to the CC-formalism becomes prohibitively expensive for heavier and larger molecules and quenchers, especially at higher collision energies with many rovibrational states involved and a large number of partial waves required for the description of the scattering process. In contrast, in QCT, the scattering process and the quantization of energy and angular momenta of the reactants are treated classically, which makes it computationally inexpensive. Unfortunately, the classical trajectory method is valid only to a certain extent. The method breaks down as collision energy decreases and fails to account for zero-point energy (in the case of vibrational transitions in molecules),[44–47] and hence cannot be adapted to study CET between the quantized states of molecules. However, one should not rule out the possibility of formulating a theoretical approach by integrating the quasi-classical trajectory treatment of scattering with the quantum treatment of rotation and/or vibration in a self-consistent way that permits energy exchange between internal and external degrees of freedom while keeping the total energy conserved.[48,49]

During the last decade we developed such mixed quantum/classical theory (MQCT) for inelastic scattering in which the relative motion of colliding partners is described using mean-field trajectories while their internal motion (rotation and vibration) is described stringently using time-dependent Schrodinger equation. This approach is expected to be more accurate than QCT but less costly than CC, especially when collision partners are heavy, and the spectrum of internal states is dense. The accuracy and feasibility of MQCT calculations was tested in a systematic study conducted over the years by comparing its results against the results of full-quantum CC-calculations. It was found that MQCT methodology gives exemplarily accurate results for diatomic and triatomic molecules,[50–55] remains computationally feasible for polyatomic molecules collided with atoms,[56,57] and even permits to embrace the complexity of molecule + molecule collisions.[42,58,59] Furthermore, MQCT can provide a distinctive time-dependent insight into the process,[55,57] that the standard time-independent quantum methods may not offer. Overall, MQCT turns out to be a powerful tool for describing CET in complex molecular systems in a broad range of collision energies.

In this paper we present the extension of MQCT methodology to describe a symmetric-top-rotor + linear-rotor collision process and report the results of first application of this theory to describe CET in $ND_3$ + $D_2$ system. Full-quantum CC-results for this process are available from recent literature[40] and can be used as a solid benchmark. Experimental studies of this process employed the technique of velocity map imaging[60] to observe correlated exchange of rotational energy between two collision partners. Excellent agreement between theoretical and experimental results was reported, which attests for the accuracy of potential energy surface (PES) and the convergence of scattering benchmark calculations.[40] Here we demonstrate that MQCT can reproduce, with good accuracy, the abovementioned correlated state-to-state transitions in two colliding molecules, which opens new opportunities for computationally efficient theoretical treatment of molecular collisions in general.

## II. THEORETICAL APPROACH

The rotations of each colliding partner is treated quantum mechanically and the wavefunction depends on the angles needed to describe individual orientations of these molecules. In general, for a symmetric-top molecule the rotations are described by a set of Euler angles $\Lambda_1 = (\alpha_1, \beta_1, \gamma_1)$. The rotational states of the molecule are quantized and are represented here by symmetrized combinations of Wigner D- functions:[61]

$$\psi_{j_1 m_1 k_1 \epsilon}(\Lambda_1) = \sqrt{\frac{2j_1+1}{8\pi^2}}\sqrt{\frac{1}{2\left(1+\delta_{k_1,0}\right)}}\left[\mathrm{D}^{j_1 *}_{k_1,m_1}(\alpha_1,\beta_1,\gamma_1) + \epsilon \mathrm{D}^{j_1 *}_{-k_1,m_1}(\alpha_1,\beta_1,\gamma_1)\right] \quad (1)$$

where $\epsilon = \pm$ is the parity. These states are labeled by quantum numbers $\{j_1 m_1 k_1 \epsilon\}$ where $j_1$ and $m_1$ represent angular momentum of the first molecule and its projection onto the axis of quantization (defined below), while $k_1$ is the projection of $j_1$ on the symmetry axis of the molecule.

The rotations of a linear rotor are described by polar angles $(\theta, \varphi)$ and its rotational eigenstates are represented by spherical harmonics $\mathrm{Y}^{j_2}_{m_2}(\theta, \varphi)$. Or, for convenience, one can use two of the three Euler angles:

$$\psi_{j_2 m_2}(\Lambda_2) = \mathrm{Y}^{j_2}_{m_2}(\beta_2, \alpha_2) \quad (2)$$

where $\Lambda_2 = (\alpha_2, \beta_2, \gamma_2 = 0)$, while $j_2$ and $m_2$ represent angular momentum of the second molecule and its projection onto the axis of quantization. Then, the coupled states of symmetric-top rotor + linear rotor can be expressed using Clebsch-Gordan (CG) coefficients $\mathrm{C}^{j,m}_{j_1 m_1, j_2 m_2}$ as follows:

$$\Psi_{nm}(\Lambda_1, \Lambda_2) = \sqrt{\frac{2j_1+1}{8\pi^2}}\sqrt{\frac{1}{2\left(1+\delta_{k_1,0}\right)}}$$

$$\times \sum_{m_1=-j_1}^{+j_1} \mathrm{C}^{j,m}_{j_1,m_1,j_2,m-m_1}\left[\mathrm{D}^{j_1 *}_{k_1,m_1}(\Lambda_1) + \epsilon \mathrm{D}^{j_1 *}_{-k_1,m_1}(\Lambda_1)\right]\mathrm{Y}^{j_2}_{m-m_1}(\Lambda_2) \quad (3)$$

Here $m$ is projection of total angular momentum $j$ of the molecule-molecule system onto the axis of quantization while $n$ is used as a composite index to label the total set of quantum numbers for the system, $n = \{j, j_1, k_1, \epsilon, j_2\}$. The CG coefficients are nonzero only if $m = m_1 + m_2$ and $|j_1 - j_2| \leq j \leq j_1 + j_2$. Thus, $(\Lambda_1, \Lambda_2)$ represents a set of quantum degrees of freedom in the system.

Time evolution of the rotational wavefunction of the system is described by expansion over a set of eigenstates:

$$\psi(\Lambda_1, \Lambda_2, t) = \sum_{nm} a_{nm}(t)\Psi_{nm}(\Lambda_1, \Lambda_2)\exp\{-\mathrm{i}E_n t\} \quad (4)$$

where $a_{nm}(t)$ is a set of time-dependent corresponding probability amplitudes, and exponential phase factors are included to simplify solution in the asymptotic range. The value of eigenstate energy $E_n$ depends on $j_1$, $k_1$ and $j_2$, but is independent of $\epsilon$ or total $j$ and $m$. Substitution of this expansion into the time-dependent Schrodinger equation and the transformation of wavefunctions into the rotating frame tied to the molecule-molecule vector $\vec{R}$ (used as a quantization axis in this body-fixed reference frame) leads to the following set of coupled equations for time-evolution of probability amplitudes:[62]

$$\dot{a}_{mn''} = -i \sum_{n'} a_{n'm} M_{n'}^{n''}(R)\, e^{i\varepsilon_{n'}^{n''} t}$$

$$-\dot{\Phi}\left[a_{n'',m-1}\sqrt{j''(j''+1) - m(m-1)} + a_{n'',m+1}\sqrt{j''(j''+1) - m(m+1)}\right]/2 \quad (5)$$

Here $\varepsilon_{n'}^{n''} = E_{n''} - E_{n'}$ is energy difference between the final and initial states of the system. Summation in the first term of this equation includes state-to-state transitions $n' \to n''$ (within each $m$) driven by real-valued, time-independent potential coupling matrix $M_{n'}^{n''}$:

$$M_{n'}^{n''}(R) = \langle \Psi_{n''m}(\Lambda_1, \Lambda_2) | V(R, \Lambda_1 \Lambda_2) | \Psi_{n'm}(\Lambda_1, \Lambda_2) \rangle \quad (6)$$

The potential energy hypersurface $V(R, \Lambda_1, \Lambda_2)$ depends on the intermolecular distance $R$ and orientation of each molecule, $\Lambda_1$ and $\Lambda_2$. The second term in Eq. (5) describes $m \pm 1 \to m$ transitions (within each $n$) due to the Coriolis coupling effect, driven by rotation of the molecule-molecule vector $\vec{R} = (R, \Phi, \Theta)$ relative to the laboratory-fixed reference frame during the course of collision.[62]

A set of spherical polar coordinates $(R, \Phi, \Theta)$ represents classical degrees of freedom in the system. They describe scattering of two collision partners relative to the laboratory-fixed reference frame and the equations for their time-evolution are obtained using Ehrenfest theorem:[62]

$$\dot{R} = \frac{P_R}{\mu} \quad (7)$$

$$\dot{\Phi} = \frac{P_\Phi}{\mu R^2} \quad (8)$$

$$\dot{P}_R = -\sum_{n'}\sum_{n''} e^{i\varepsilon_{n'}^{n''}t} \sum_m \frac{\partial M_{n'}^{n''}}{\partial R} a_{n''m}^* a_{n'm} + \frac{P_\Phi^2}{\mu R^3} \tag{9}$$

$$\begin{aligned}\dot{P}_\Phi = -i\sum_{n'}\sum_{n''} e^{i\varepsilon_{n'}^{n''}t} \sum_m M_{n'}^{n''} \\ \times \Big[a_{n''m-1}^* a_{n'm}\sqrt{j''(j''+1)-m(m-1)} \\ + a_{n''m+1}^* a_{n'm}\sqrt{j''(j''+1)-m(m+1)} \\ - a_{n''m}^* a_{n'm-1}\sqrt{j'(j'+1)-m(m-1)} \\ - a_{n''m}^* a_{n'm+1}\sqrt{j'(j'+1)-m(m+1)}\Big]/2\end{aligned} \tag{10}$$

It appears that only the equations for $R$, $\Phi$ and their conjugate momenta $P_R$, $P_\Phi$ are needed. Since the trajectory is planar, one can restrict consideration to the equatorial plane $\Theta = \pi/2$ with $\dot{\Theta} = 0$. [52,62] Note that classical orbital angular momentum $\dot{\Phi}(t)$ drives Coriolis transitions in the quantum equation of motion, Eq. (5), while the quantum probability amplitudes $a_{nm}(t)$ create a mean-field potential in the classical equations of motion, Eqs. (9-10), providing a link between quantum and classical degrees of freedom. It was demonstrated that the total energy, which is the sum of rotational (quantum) and translational (classical), is conserved along these mixed quantum/classical trajectories.[52,62]

In principle, matrix elements of Eq. (6) can be computed by a four-dimensional numerical quadrature:

$$\begin{aligned}M_{n'}^{n''}(R) = 2\pi \int_0^\pi \sin\beta_1\, d\beta_1 \int_0^{2\pi} d\gamma_1 \\ \times \int_0^{2\pi} d\alpha_2 \int_0^\pi \sin\beta_2\, d\beta_2\, V(R,\Lambda_1\Lambda_2)\Psi_{n''}^*(\Lambda_1\Lambda_2)\Psi_{n'}(\Lambda_1\Lambda_2)\end{aligned} \tag{11}$$

The factor of $2\pi$ come from the analytical integration over $\alpha_1$. This can be done because potential energy of the system depends only on the relative orientations of two molecules, given by the difference $\alpha_2 - \alpha_1$. One can set $\alpha_1 = 0$, which makes $V(R,\Lambda_1\Lambda_2)$ independent of $\alpha_1$. In practice, the multi-dimensional quadrature is numerically expensive. It is better to expand $V(R,\Lambda_1\Lambda_2)$ over a set of suitable angular functions $\tau_{\lambda_1\mu_1\lambda_2\lambda}(\beta_1,\gamma_1,\alpha_2,\beta_2)$ with $R$-dependent expansion coefficients $\upsilon_{\lambda_1\mu_1\lambda_2\lambda}(R)$ obtained by projecting $V$ onto these $\tau$-functions at each value of $R$ within a predefined grid. These projections are also computed by numerical quadrature, but the number of expansion functions is much smaller than the number of individual matrix elements. Then, at these values of $R$, the potential can be represented analytically:

$$V(R,\Lambda_1\Lambda_2) = \sum_{\lambda_1\mu_1\lambda_2\lambda} \upsilon_{\lambda_1\mu_1\lambda_2\lambda}(R)\tau_{\lambda_1\mu_1\lambda_2\lambda}(\beta_1,\gamma_1,\alpha_2,\beta_2) \tag{12}$$

For a symmetric-top rotor + linear rotor system a suitable set of functions is given by:[42]

$$\begin{aligned}\tau_{\lambda_1\mu_1\lambda_2\lambda}&(\beta_1,\gamma_1,\alpha_2,\beta_2)\\ &= \sqrt{\frac{2\lambda_1+1}{4\pi}} \sum_{\eta=-\min(\lambda_1,\lambda_2)}^{+\min(\lambda_1,\lambda_2)} \mathrm{C}_{\lambda_1,\eta,\lambda_2,-\eta}^{\lambda,0}\\ &\times \left[\mathrm{D}_{\mu_1,\eta}^{\lambda_1 *}(0,\beta_1,\gamma_1) + (-1)^{\lambda_1+\mu_1+\lambda_2+\lambda}\mathrm{D}_{-\mu_1,\eta}^{\lambda_1 *}(0,\beta_1,\gamma_1)\right]\mathrm{Y}_{-\eta}^{\lambda_2}(\beta_2,\alpha_2)\end{aligned} \tag{13}$$

which uses spherical harmonics $\mathrm{Y}_{\eta}^{\lambda}$, Wigner D-functions and CG coefficients (see above). The meaning of indexes $\lambda_1$ ($\mu_1$), $\lambda_2$ and $\lambda$ are analogues to angular momenta for the molecule one (its projection onto symmetry axis), the molecule two, and the entire system, respectively. Substitution of Eq. (13) into Eq. (12), and then into Eq. (11), and a somewhat lengthy derivation outlined in the *Supplemental Information* (SI), gives the following final expression:

$$\begin{aligned}M_{n'}^{n''}(R) = &\sqrt{\frac{2j_1'+1}{2j_1''+1}}\sqrt{\frac{2j_2'+1}{2j_2''+1}}\frac{1}{\sqrt{2\left(1+\delta_{k_1',0}\right)2\left(1+\delta_{k_1'',0}\right)}}\sum_{\lambda_1\mu_1\lambda_2\lambda}\upsilon_{\lambda_1\mu_1\lambda_2\lambda}(R)\sqrt{\frac{2\lambda_1+1}{4\pi}}\sqrt{\frac{2\lambda_2+1}{4\pi}}\\ &\times \mathrm{C}_{j_2',0,\lambda_2,0}^{j_2'',0}\sum_{m_1'=-j_1'}^{+j_1'}\mathrm{C}_{j_1',m_1',j_2',m-m_1'}^{j',m}\sum_{\eta=-\mathrm{mi}\ (\lambda_1,\lambda_2)}^{+\min(\lambda_1,\lambda_2)}\mathrm{C}_{j_1'',m_1'-\eta,j_2'',m-(m_1'-\eta)}^{j'',m}\,\mathrm{C}_{\lambda_1,\eta,\lambda_2,-\eta}^{\lambda,0}\mathrm{C}_{j_1',m_1'-\eta,\lambda_1,\eta}^{j_1'',m_1'}\\ &\times \mathrm{C}_{j_2',m-(m_1'-\eta),\lambda_2,-\eta}^{j_2'',m-m_1'}\begin{bmatrix} \left(C_{j_1',k_1',\lambda_1,\mu_1}^{j_1'',k_1''} + (-1)^{\lambda_1+\mu_1+\lambda_2+\lambda}C_{j_1',k_1',\lambda_1,-\mu_1}^{j_1'',k_1''}\right)\\ +\ \epsilon'\left(C_{j_1',-k_1',\lambda_1,\mu_1}^{j_1'',k_1''} + (-1)^{\lambda_1+\mu_1+\lambda_2+\lambda}C_{j_1',-k_1',\lambda_1,-\mu_1}^{j_1'',k_1''}\right)\\ +\ \epsilon''\left(C_{j_1',k_1',\lambda_1,\mu_1}^{j_1'',-k_1''} + (-1)^{\lambda_1+\mu_1+\lambda_2+\lambda}C_{j_1',k_1',\lambda_1,-\mu_1}^{j_1'',-k_1''}\right)\\ +\epsilon'\epsilon''\left(C_{j_1',-k_1',\lambda_1,\mu_1}^{j_1'',-k_1''} + (-1)^{\lambda_1+\mu_1+\lambda_2+\lambda}C_{j_1',-k_1',\lambda_1,-\mu_1}^{j_1'',-k_1''}\right)\end{bmatrix}\end{aligned} \tag{14}$$

In order to test the accuracy of PES expansion method, we computed the values of matrix elements for a small subset of states that include combinations of ground and excited states of both collisions partners ($j_{1k_1}^{\pm} j_2 = 1_1^+0,\ 1_1^-0,\ 2_1^+0,\ 2_1^-0,\ 3_1^+0,\ 3_1^-0,\ 1_1^+2,\ 1_1^-2,\ 2_1^+2,\ 2_1^-2,\ 3_1^+2,\ 3_1^-2$) at two values of the molecule-molecule distance, $R$ = 6.8 and 6.9 Bohr. In these composite state labels, the rotational levels of $ND_3$ are labeled as $j_{1k_1}^{\pm}$, followed by $j_2$ of $D_2$. Expansion terms with $\lambda_1$ and $\mu_1$ up to 6, $\lambda_2$ and $\lambda$ up to 4 and 10, respectively, were included in the analytic representation of the PES. Numerical quadrature was used to obtain, by projection, the expansion coefficients $v_{\lambda_1\mu_1\lambda_2\lambda}(R)$ needed in Eq. (14), and to calculate the same matrix elements by direct integration using Eq. (11). Equidistant grids with 40 points were used for $\alpha$ and $\gamma$, and a Gauss-Legendre method with 20 points was used for $\beta$. The values of computed matrix elements in this test were found to be in the range from $10^{-4}$ to $10^{2}$ $cm^{-1}$, but the difference between two methods of calculations was found to be on the order of $10^{-10}$ $cm^{-1}$, from which we conclude that both methods work as expected. Direct integration of Eq. (11) helps to ensure that the analytical expansion of Eq. (14) is correct. The numerical speed-up of computing matrix elements by expansion is very significant, by a factor of ~ 11 for the subset of states indicated above.

## III. RESULTS AND DISCUSSION

Within MQCT framework, we computed state-to-state transition cross sections between the rotational states of $ND_3$ + $D_2$ system. Following Gao et al [40] we adopted a basis set that includes the rotational states of $ND_3$ up to $j_1 = 8$ with $k_1 \leq j_1$ restricted to $k_1 = 1, 2, 4, 5, 7, 8$ (sometimes called "para" $ND_3$) and ortho-$D_2$ with $j_2 = 0, 2, 4$, which led to 138 nondegenerate quantum states of the molecule-molecule system. Energies of these states cover the range up to $E \sim 900$ $cm^{-1}$. Considering all possible values of total angular momentum (up to $j = 12$) and its projection (in the range $-12 \leq m \leq 12$) resulted in 7,770 quantum states overall and 1,903,543 state-to-state transitions with non-zero matrix elements $M_{n'}^{n''}$. Out of these transitions, 1,412,776 have matrix elements above $10^{-4}$ $cm^{-1}$ so, in principle, this matrix could be truncated to simplify calculations (was not implemented here because the calculations were affordable anyway). The values of matrix elements were computed on a grid of 86 points in the range $3.5 \leq R \leq 25$ Bohr and interpolated using cubic spline. We carefully examined the convergence of our calculations with respect to several input parameters in the MQCT code. To ensure that symmetry properties were incorporated and taken into consideration correctly we also tried to include all states of $ND_3$ (both para and ortho states with $k_1 = 0, 3, 6$) and computed the state-to-state transition matrix elements. In this test we found that for all ortho-para transitions the matrix elements were numeral zeros. For simulations of collision dynamics, we started MQCT trajectories at $R_{\max} = 25$ Bohr with impact parameters up to $b_{\max} = 15$ Bohr and propagated the equations of motion using 4th order Runge-Kutta method with a step size of $\Delta t = 50$ a.u. ~ 1.2 fs. The relative error of total energy conservation in Eq. (7-10) and of the wavefunction norm conservation in Eq. (5) were within 0.01% of the initial values.

Kinetic energy of $ND_3 + D_2$ collision was varied through a broad range from 1 cm$^{-1}$ to $10^4$ cm$^{-1}$. Figure 1 gives examples of energy dependence of cross section for several state-to-state transitions. Several more transitions are included in Fig. S1 of SI. For each transition we plotted the results of "direct" MQCT calculations of quenching and excitation cross sections (solid lines), but also the results of "reverse" approach (dashed lines), where the value of excitation cross section is derived from computed quenching cross section, and vice versa, using the principle of microscopic reversibility:[53]

$$(2j_1 + 1)(2j_2 + 1)E\sigma_{j_1 j_2 \to j_1' j_2'} = (2j_1' + 1)(2j_2' + 1)E'\sigma_{j_1' j_2' \to j_1 j_2} \quad (15)$$

Here $E$ and $E'$ correspond to collision energies for quenching and excitation processes, respectively, of the same transition $(j_1 j_2) \leftrightarrow (j_1' j_2')$ at the same total energy, so that $E = E' + \Delta E$, where $\Delta E$ is the absolute value of energy difference between states $(j_1 j_2)$ and $(j_1' j_2')$. The difference of cross sections computed using direct and reverse methods can be used to estimate the accuracy of MQCT. The data presented in Fig. 1 and Fig. S1 demonstrate that the microscopic reversibility is generally satisfied in the case of MQCT calculations for all transitions in $ND_3 + D_2$ system. For each excitation process presented in Fig. 1 and Fig. S1 the energy dependence of cross section exhibits a very clear threshold behavior (black curves). Namely, as collision energy is reduced and becomes merely sufficient to excite the system, the value of cross section for the process quickly vanishes. In contrast, for all quenching processes (red curves in Fig. 1 and Fig. S1) the value of cross section shows a steady growth towards low collision energies. One can see that at high collision energies the principle of microscopic reversibility is accurately satisfied (solid and dashed lines coincide), but some deviations can be noticed near thresholds for excitation transitions, and at very low collision energies for quenching transitions. This is expected because an approximate trajectory-based method may become less accurate at low energies.

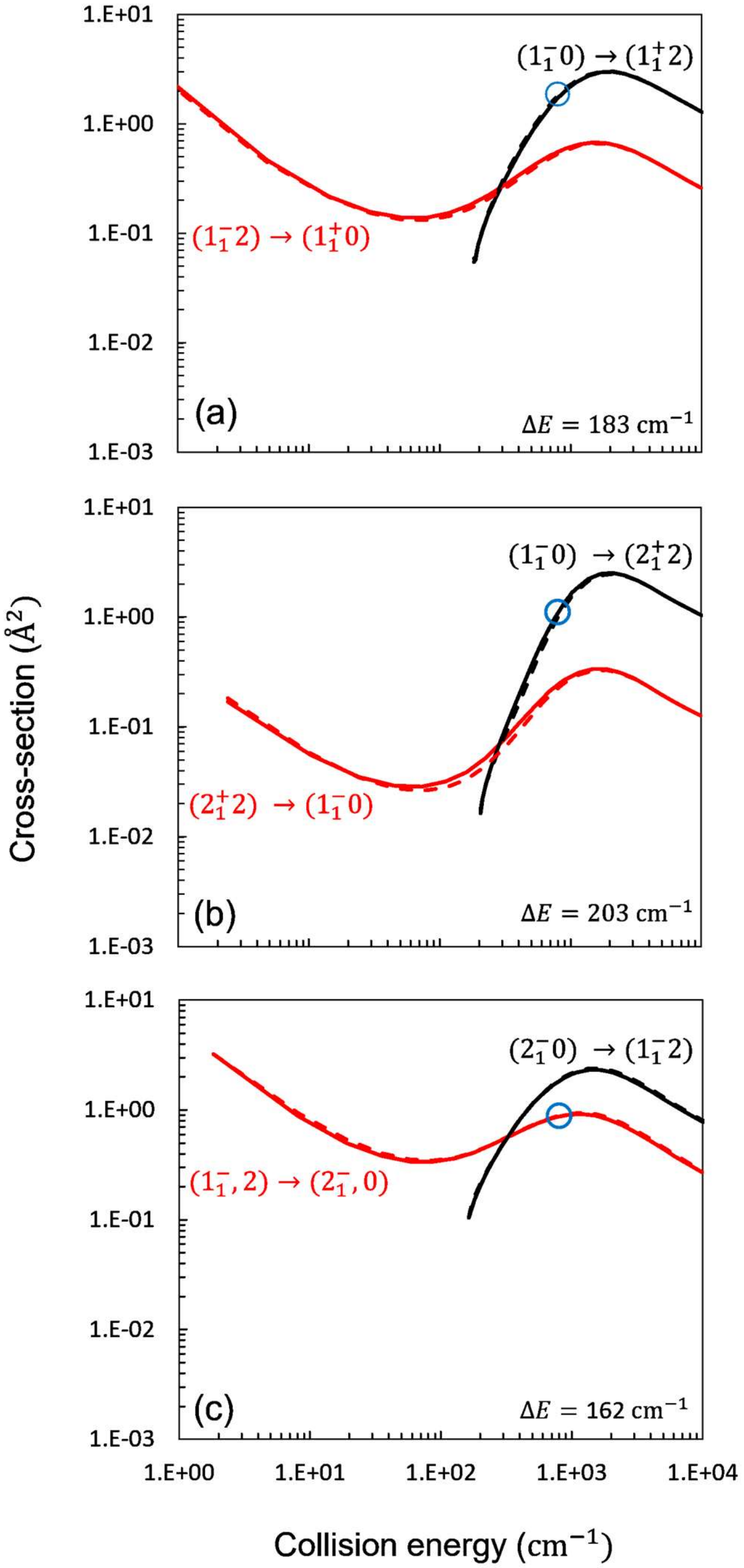

**Figure 1:** The test of microscopic reversibility for transitions between several rotational states of $ND_3 + D_2$ system, labeled as $(j_{1}{}_{k}^{\pm} j_2)$. Cross sections are plotted as a function of collision energy. The data obtained by "direct" MQCT calculations are shown by solid lines, whereas dashed lines represent the results of "reverse" calculations. Red color is used for quenching processes, black color is for excitation processes. Blue symbol indicates full-quantum results of Ref. [40] The value of energy difference is given for each transition.

It should be emphasized that the three transitions shown in Fig. 1 belong to three different types of processes. Namely, in Fig. 1a we deal with a nearly elastic parity-changing transition $1_1^- \leftrightarrow 1_1^+$ in $ND_3$ that occurs simultaneously with excitation or quenching of $D_2$, $0 \leftrightarrow 2$. In Fig. 1b both $ND_3$ and $D_2$ undergo the same process: either both are being excited, or both are being quenched. Finally, in Fig. 1c we see excitation of $ND_3$ simultaneously with quenching of $D_2$, and vice versa. Fifteen transitions presented in Fig. S1 of SI are also split into these three groups. Based on this large set of data we can conclude that microscopic reversibility is satisfied reasonably well in the results of MQCT calculations for a variety of energy exchange processes and in a broad range of collision energies, except at the lowest energies where some differences are observed for some transitions (see Fig. S1).

Also, from the data presented in the fifteen frames of Fig. S1 we can derive the following trend: at high collision energy the difference between excitation and quenching cross sections (the spacing between black and red curves within each frame) correlate with energy difference between the initial and final states of the $ND_3 + D_2$ system. The smallest difference between the excitation and quenching cross sections is found for $4_1^{\pm}0 \leftrightarrow 1_1^{-}2$ transition that have the smallest energy difference of 90 cm$^{-1}$. The largest difference of cross sections is found for $4_1^{\pm}2 \leftrightarrow 1_1^{-}0$ transition that have the largest energy difference of 275 cm$^{-1}$. Other transitions fall in between and follow this interesting trend. Blue symbols included in Fig. 1 and Fig. S1 give the values of transition cross section from literature, computed using full-quantum CC method[[40]] at collision energy of 800 cm$^{-1}$, and we see that these data are in good agreement with our MQCT results.

In order to obtain more insight into the properties of different state-to-state transition processes in $ND_3 + D_2$ collisions, we plotted, for all of transitions discussed above, the dependence of transition probability on collision impact parameter, sometimes called the opacity functions. The impact parameter $b$ is related to the orbital angular momentum quantum number $\ell$ through the following relationship $\ell(\ell+1) = k^2b^2$ where the magnitude of wave vector $k$ is determined by collision energy, $\hbar k = P = \sqrt{2\mu E}$. For example, at collision energy of 800 cm$^{-1}$ the value of $b \sim 12$ Bohr corresponds to $\ell = 84$. Figure 2 gives examples for three transitions out of which, again, one changes the parity of $ND_3$ (while $D_2$ is excited of quenched, Fig. 2a and below), second is a transition in which two collision partners undergo the same kind of change (both are either excited or quenched, Fig. 2b and below), and third is a transition where two collision partners exchange energy (one is excited while the other is quenched, Fig. 2c and below).

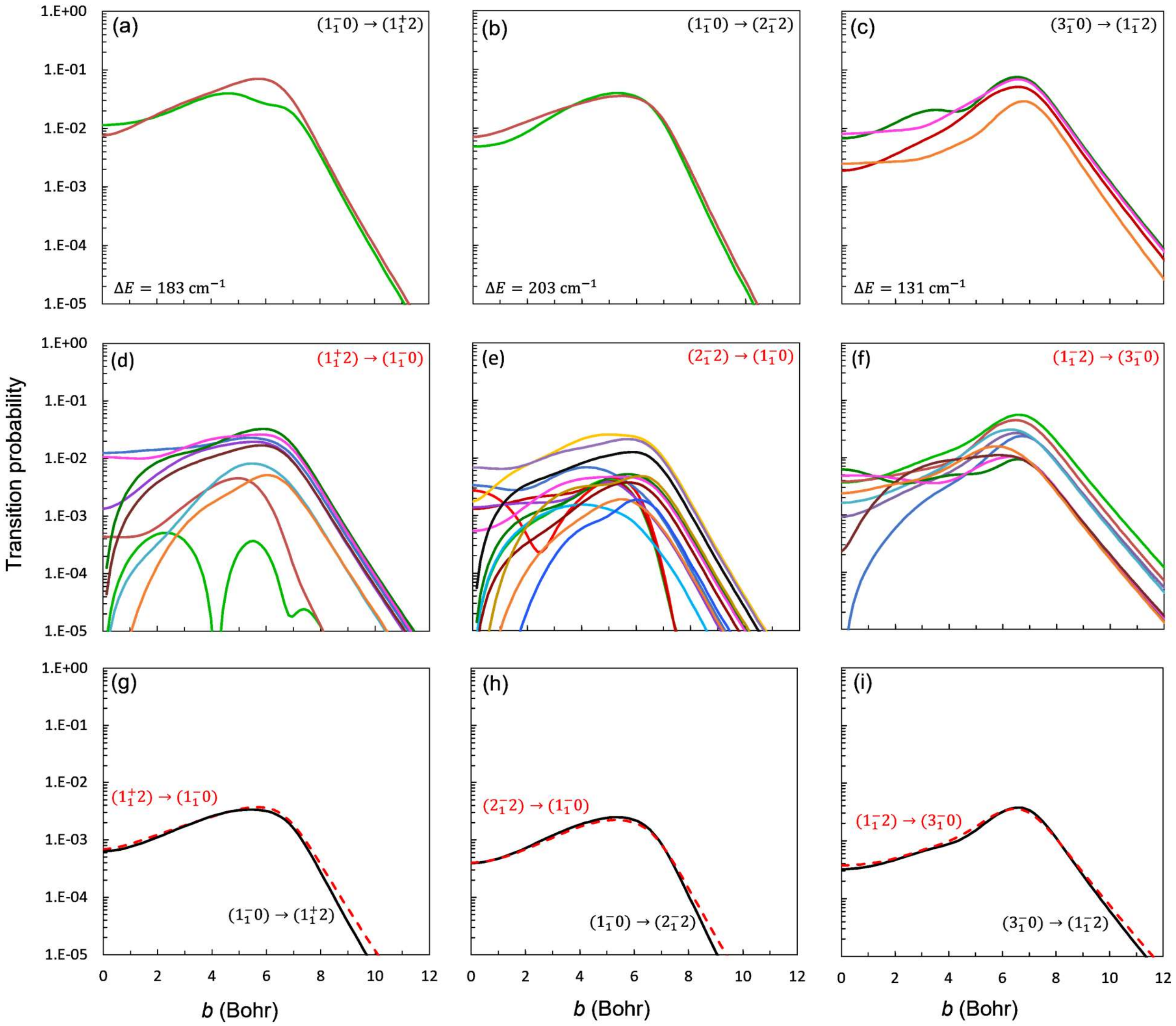

**Figure 2:** Opacity functions for several transitions between the rotational states of $ND_3$ + $D_2$ system, labeled as $(j_{1}{}_{k}^{\pm}j_2)$. Transition probabilities are plotted as a function of collision impact parameter. Collision energy is 800 cm$^{-1}$. In the upper two rows of frames, the opacity functions are given for all individual $(j, m)$-components of the initial state. In the lower row of frames, the opacity functions are averaged over the initial $(j, m)$-states, divided by degeneracy of the final state, and the results for excitation and quenching are plotted together, in order to check microscopic reversibility.

In the upper two rows of frames in Fig. 2 different lines within the same frame correspond to different possible values of the total $j$ and $m$ in the initial state. For example, in the process $3_1^-0 \rightarrow 1_1^-2$ of Fig. 2c with the initial $j_1 = 3$ and $j_2 = 0$, only one initial value of total angular momentum is possible, $j = 3$, that comes with $|m| = 0, 1, 2$ and 3. This gives four different initial $(j, m)$-states and four curves plotted in Fig. 2c that correspond to these initial conditions. Negative values of $m$ give the same transition probabilities as positive once, and thus are not plotted (but they are counted in the reversibility principle, see below). However, for the reverse process $1_1^-2 \rightarrow 3_1^-0$ of Fig. 2f with the initial $j_1 = 1$ and $j_2 = 2$, three different values of total angular momentum are possible: $j = 1, 2$ and 3 that come with two, three and four values of non-negative $m$, respectively, which gives nine different initial $(j, m)$-states. So, nine curves plotted in Fig. 2f correspond to these initial states, and so on.

From Fig. 2 one can see that although all opacity functions exhibit maximum in the vicinity of the impact parameter $b \sim 6$ Bohr, it appears that, even within the same state-to-state transition, the behaviors of opacity functions for different initial $(j, m)$-states are quite different in the range of smaller impact parameters. For example, for $1_1^-2 \rightarrow 3_1^-0$ transition discussed above, one opacity function (blue in Fig. 2f) vanishes in the $b \rightarrow 0$ limit, while eight others remain finite at $b = 0$. For $1_1^+2 \rightarrow 1_1^-0$ transition, five opacity functions vanish in the $b \rightarrow 0$ limit (orange, blue, green, brown, turquoise in Fig. 2d), while four others remain finite at $b = 0$. Finally, for $2_1^-2 \rightarrow 1_1^-0$ transition, eight opacity functions (blue, orange, brown, maroon, turquoise, green, black in Fig. 2e) vanish in the $b \rightarrow 0$ limit, while seven others remain finite at $b = 0$. These two scenarios correspond to Coriolis-driven vs potential-driven transition processes, respectively. It appears that, in most cases, for the same transition (say $1_1^-2 \rightarrow 3_1^-0$), matrix elements $M_{n'}^{n''}$ are non-zero only for some values of the total $j$ and $m$, but turn to zero for several other values of total $j$ and $m$. In those cases when matrix elements $M_{n'}^{n''}$ are zero, the state-to-state transitions still proceed but through the Coriolis coupling, driven by second term in Eq. (5). Overall, the effect of Coriolis coupling is often smaller than the potential coupling, but from transition probabilities presented in Fig. 2 one can see that in general it is not negligible and must be properly taken into account. Table S1 in SI lists potential-driven and Coriolis-driven $(j, m)$-components for all transitions discussed in this work.

Equation (15) above represents the principle of microscopic reversibility in terms of cross sections, but it is also instructive to write it in terms of transition probabilities of the individual MQCT trajectories, like those presented in Fig. 2. For this, it is important to recall that cross sections in Eq. (15) represent the sum over final and average over initial $(j, m)$-states within each transition. If $p$ represents the corresponding transition probability (summed over final and averaged over initial states), then the principle of microscopic reversibility can be rewritten as:

$$\frac{p_{j_1 j_2 \to j_1' j_2'}}{(2j_1'+1)(2j_2'+1)} = \frac{p_{j_1' j_2' \to j_1 j_2}}{(2j_1+1)(2j_2+1)} \tag{16}$$

Probabilities in the left- and right-hand sides of Eq. (16) should be obtained for the same collision impact parameter and at the same energy $U$ of MQCT trajectory. Note that collision energies $E$ and $E'$ for quenching and excitation processes that appear in Eq. (15), are not anymore present in Eq. (16). They analytically cancel because cross sections are inversely proportional to collision energy and proportional to transition probability, $\sigma \sim p/k^2$. In this form, left- and right-hand sides of Eq. (16) represent probability per one $(j, m)$-state on average, i.e., averaged over both final and initial states within a transition. To guarantee reversibility, these probabilities should be equal for direct and reverse processes.

In the lower row of Fig. 2 we present these probabilities for comparison, for three transitions. Black curves were computed from the data presented in the 1$^{st}$-row of frames in Fig. 2, while red curves were computed from the data presented in 2$^{nd}$-row of frames in Fig. 2 (i.e., for the corresponding reverse transitions). Their comparison, presented in the 3$^{rd}$-row of Fig. 2, indicates a very good agreement between the two, which means that MQCT satisfies reversibility even at the level of individual trajectories. This is not immediately obvious from visual comparison of the corresponding frames in the 1$^{st}$ and 2$^{nd}$ rows of Fig. 2, because they contain a different number of curves due to different number of the initial $(j, m)$-states, and, moreover, some of these curves correspond to the potential driven transitions, while others are Coriolis driven and the number of processes of each type changes from one transition to another, as one can see in Fig. 2. Still, the microscopic reversibility is satisfied in all cases! In SI, we plotted the left- and right-hand sides of Eq. (16) for fifteen transitions in $ND_3$ + $D_2$ system in Fig. S2, and for fourteen more transitions in Fig. S3. In all these cases we observed good systematic agreement with the principle of microscopic reversibility (see SI).

In order to obtain time-dependent insight into the process of energy exchange between two collision partners, we plotted time-evolution of state populations along the individual MQCT trajectories with chosen values of impact parameter $b$, indicated by small green arrows in Fig. S2 and S3. In Fig. 3 we present one example with collision impact parameter $b = 5.58$ Bohr for the initial state $1_1^-0$ (the case of total $j = 1$ and $m = 0$ is shown, other cases look similar). Referring to Fig. S2 of SI one can see that the maximum values of opacity functions for several transitions are found near this value of impact parameter, which means that the collision event presented in Fig. 3 is typical and makes a substantial contribution to the total cross section. The typical time of trajectory propagation is $\sim 1$ ps, but the closest approach of two collision partners (the turning point of trajectory) happens at $\sim 0.56$ ps, which represents the midpoint of collision event, indicated in Fig. 3 by vertical dashed line.

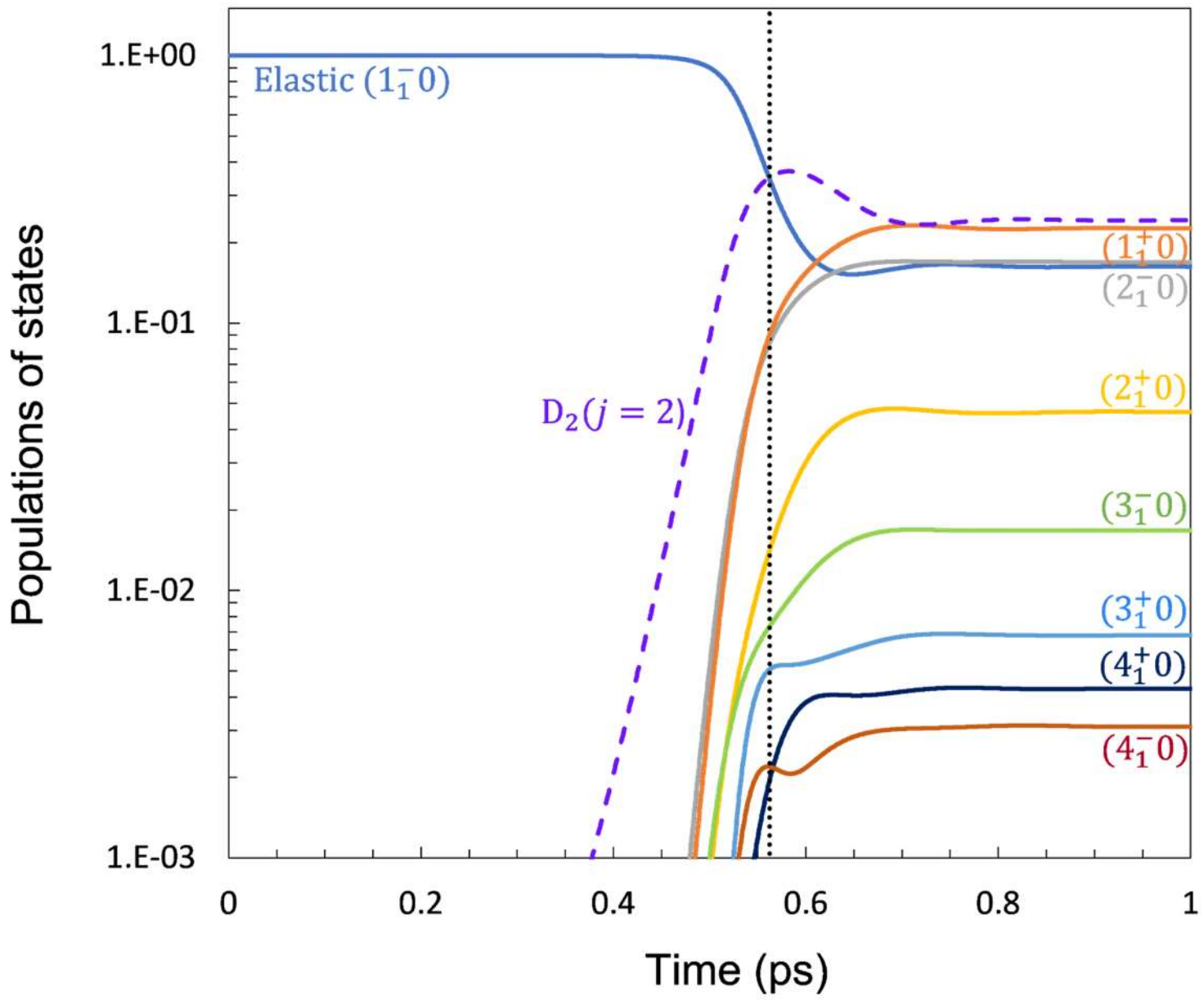

**Figure 3:** Evolution of state populations in $ND_3$ and $D_2$ along one typical MQCT trajectory at collision energy of 800 $cm^{-1}$. The impact parameter is $b = 5.58$ Bohr, which corresponds to the orbital angular momentum quantum number $\ell = 35$. The initial states are $ND_3$ $(1_1^-)$ and $D_2$ $(j = 0)$. The total population of $D_2$ $(j = 2)$ is shown by dashed purple line. The final $ND_3 + D_2$ $(j = 0)$ states are labeled and are indicated by color. A vertical dashed line indicates the moment of closest approach of two molecules.

Analysis of Fig. 3 permits us to tell an interesting story of energy transfer in the process of $ND_3$ excitation by collision with $D_2$. One can see that population of the initial state $ND_3(1_1^-)$ drops significantly during the collision event and this happens rather quickly, within $\sim 0.10$ ps. Over 80% of the initial state population is transferred to several excited $ND_3$ states, but this does not happen directly. Indeed, in Fig. 3 one can notice that populations of the excited $ND_3$ states $1_1^+$, $2_1^-$, $2_1^+$ and $3_1^-$ grow monotonically and slowly, reaching their asymptotic values on a longer time scale, $\sim 0.20$ ps. During this time the excited state of $D_2(j = 2)$ acts as an effective intermediate state in the process of energy transfer from the translational energy of collision partners to the excited states of $ND_3$. Namely, the excited state of $D_2$ starts receiving its population early, $\sim 0.10$ ps before it goes to the excited states of $ND_3$. Note that in the middle of collision event, when the population of $D_2(j = 2)$ becomes equal to the population of the elastic channel $1_1^-0$, the populations of excited states of $ND_3$ remain small, under 10%. The maximum population of $D_2(j = 2)$ reaches 37% but then drops to about 23%, releasing a significant portion of its population to the excited states of $ND_3$.

We think that the reason for the efficiency of this two-step energy transfer mechanism lies in the energy scales in the system. Namely, the excitation energy of $D_2(j = 2)$ is 183 $cm^{-1}$, while the energy spacings between the rotational states of $ND_3$ are much smaller: 20.57, 30.85 and 41.15 $cm^{-1}$ between the consecutive states. The former is much closer to the kinetic energy of collision, which is 800 $cm^{-1}$. Therefore, the rotational states of $D_2$ represent a more suitable sink for a large translational energy of collision partners. The rotational states of $ND_3$ receive energy from the excited $D_2$, rather than from the translational motion directly, again, due to similar energy scales.

In Figure 4 we present another example of evolution of individual state populations along MQCT trajectory, for a different initial state $1_1^-2$. This case permits us to see what happens when a significant amount of energy is already stored in $D_2(j = 2)$ prior to the $ND_3$ + $D_2$ collision. The values of total $j = 1$ and $m = 0$ were chosen for the initial state, as before. The value of impact parameter was $b = 5.00$ Bohr and the moment of closest approach of two collision partners was at $\sim 0.47$ ps. From Fig. 4 we see that in this case the populations of excited $ND_3$ states grow monotonically within both the excited $D_2(j = 2)$ and ground $D_2(j = 0)$ manifolds, with the excited state being ahead of the ground, both in terms of the beginning of the excitation process (that starts about $\sim 0.05$ ps earlier for the excited $D_2$) and in terms of the final transition probability (that is an order of magnitude larger for the excited $D_2$). We conclude that if the initial state of $D_2$ is already excited $(j = 2)$, the process of $ND_3$ excitation is direct, in contrast to the two-step process discussed above for the case of ground state $D_2(j = 0)$.

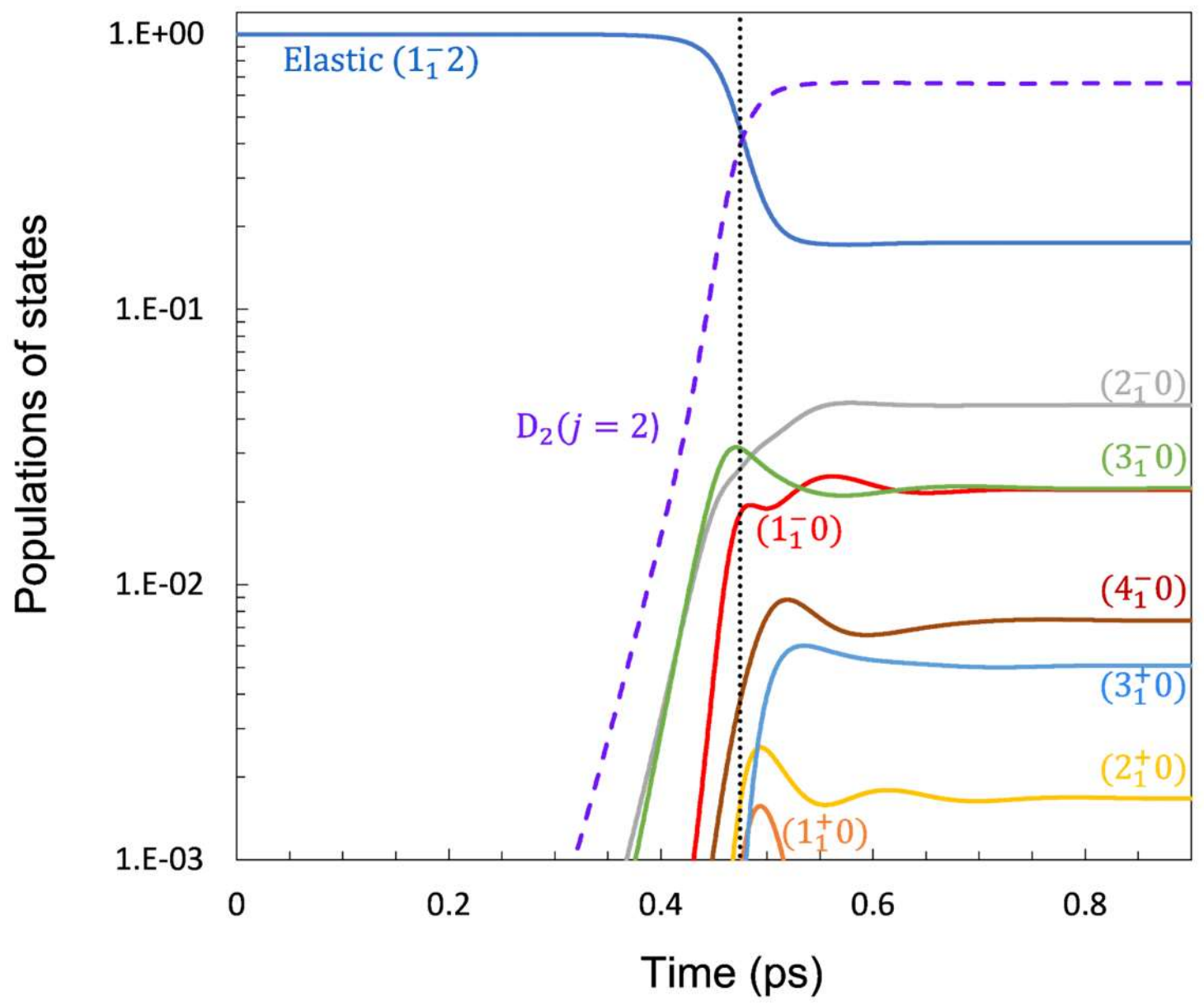

**Figure 4:** Evolution of state populations in $ND_3$ and $D_2$ along one typical MQCT trajectory at collision energy of 800 $cm^{-1}$. The impact parameter is $b = 5.0$ Bohr, which corresponds to the orbital angular momentum quantum number $\ell = 35$. The initial states are $ND_3(1_1^-)$ and $D_2(j = 2)$. The total population of $D_2(j = 2)$ except elastic $(1_1^-2)$ is shown by dashed purple line. The final $ND_3 + D_2(j = 0)$ states are labeled and are indicated by color. A vertical dashed line indicates the moment of closest approach of two molecules.

One may also notice from Fig. 4 that in this example the populations of all positive parity states are much smaller compared to their negative parity counterparts. For example, the population of $1_1^+0$ remains small during the collision event and vanishes after the collision. The population of state $4_1^+0$ remains below $10^{-3}$, outside of the frame in Fig. 4. This is very different from the case presented in Fig. 3, where cross sections for $ND_3$ states of two parities are comparable. We found that for these initial conditions (namely, total $j = 1$ and $m = 0$ of $1_1^-2$-state) the parity changing transitions to $D_2$ $(j = 0)$ are Coriolis-driven. For these processes the potential coupling matrix elements $M_{n'}^{n''}$ of Eq. (6) are zero due to symmetry. The time evolution of corresponding probability amplitudes is driven only by the second term in Eq. (5) -- the Coriolis coupling term, since the contribution of the first term (the potential coupling term) is null. It should not be mistakenly concluded, though, that all parity changing transitions are exclusively Coriolis driven. This is not the case. For example, the initial state with total $j = 1$ and $m = 1$ (of the same $1_1^-2$-channel) has non-zero potential coupling matrix elements $M_{n'}^{n''}$ that would make a significant contribution to parity changing transitions. As mentioned above, for about a half of the initial $(j, m)$-states the potential coupling matrix elements are zero, and in those case the Coriolis coupling is the only mechanism.

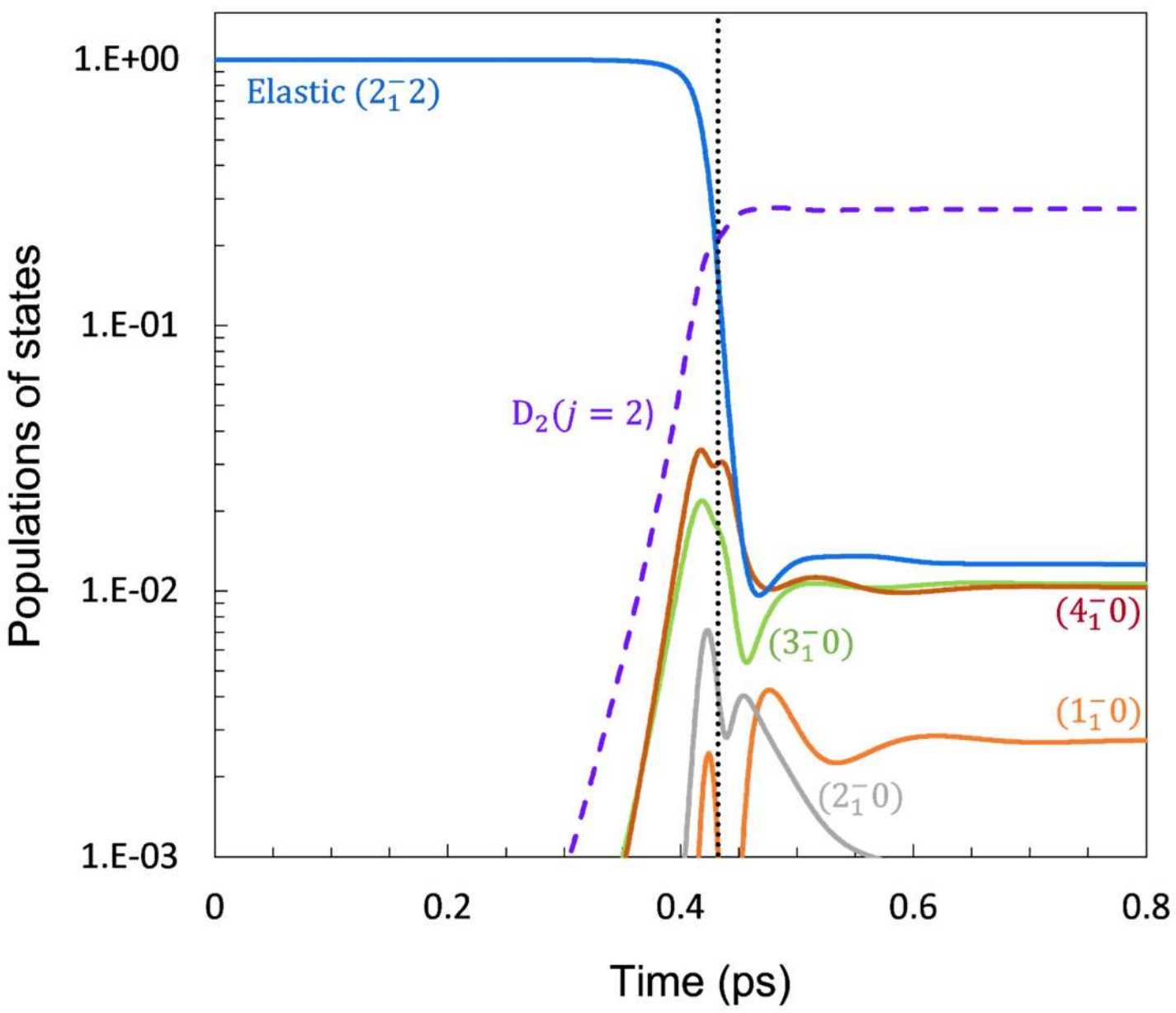

**Figure 5:** Evolution of state populations in $ND_3$ and $D_2$ along one typical MQCT trajectory at collision energy of 800 $cm^{-1}$. The impact parameter is $b = 0$, which corresponds to the orbital angular momentum quantum number $\ell = 0$. The initial states are $ND_3$ $(2_1^-)$ and $D_2$ $(j = 2)$. The total population of $D_2(j = 2)$ except elastic $(2_1^-2)$ is shown by dashed purple line. The final $ND_3$ + $D_2(j = 0)$ states are labeled and are indicated by color. A vertical dashed line indicates the moment of closest approach of two molecules.

In Figure 5 we present one more example of MQCT trajectory, for the initial state $2_1^-2$ (total $j$=0 and $m = 0$) and the impact parameter $b = 0$. This trajectory corresponds to a head on collision with orbital angular momentum $\ell = 0$. The initial value of angular speed $\dot{\Phi}$ of $\vec{R}$-vector rotation is also zero, and it remains zero during the entire collision event. Collision partners approach each other along a straight line, reach the turning point and scatter back to the asymptotic region, all along a straight-line trajectory. In this case the contribution of Coriolis coupling is null because the corresponding term in Eq. (5) has a pre-factor of $\dot{\Phi}$. Indeed, the Coriolis coupling between the quantum states in MQCT method is driven by rotation of the molecule-molecule vector $\vec{R}$ treated classically. If $\dot{\Phi} = 0$ the vector $\vec{R}$ does not rotate at all, there is no Coriolis coupling and no Coriolis-driven state-to-state transitions. This is exactly the case presented in Fig. 5. Therefore, for this special trajectory, there are no transitions into the states of positive parity.

One of the main goals of this paper was to compare our MQCT results with full-quantum results of Gao *et al.* [40] It should be mentioned, though, that the labelling of rotationally degenerate states in their work is different. Namely, they label the states of $ND_3$ using umbrella inversion symmetry $s$, related to our parity $\epsilon$ through the following expression: $s = (-\epsilon)(-1)^{j_1}$, as emphasized on Page 2 of Ref. [31] (first line of right column). Therefore, the initial state labeled as $1_1^-$ is the same state both here and there, and this is true for all rotational states with odd values of $j_1$. But, for the states with even values of $j_1$ our parity is opposite to their inversion symmetry. In what follows we keep using our notations, even when we use their data for comparison.

Figure 6 represents the comparison of our state-to-state transition cross sections (obtained using MQCT) against the full-quantum data from Fig. 12 and 13 of Ref.[40] Upper frame gives excitation of $ND_3(1_1^-)$ in coincidence with excitation of case of $D_2(j = 0)$, while lower frame gives excitation of $ND_3(1_1^-)$ in coincidence with quenching of case of $D_2(j = 2)$. One important conclusion is that MQCT gives a reliable description of the process including all kinds of transitions and energy exchange pathways between collision partners. On a quantitative level, MQCT seem to underestimate the values of cross sections by about 8% (RMSD for a set of 16 transitions in Fig. 6), which is quite satisfactory, considering the approximate quantum/classical nature of the method. It is interesting to note that in the case of $D_2$ excitation, MQCT shows a larger deviation from full quantum results (11% RMSD for 8 transitions in the upper frame of Fig. 6) compared to the case of $D_2$ quenching (2% RMSD for 8 transitions in the lower frame of Fig. 6).

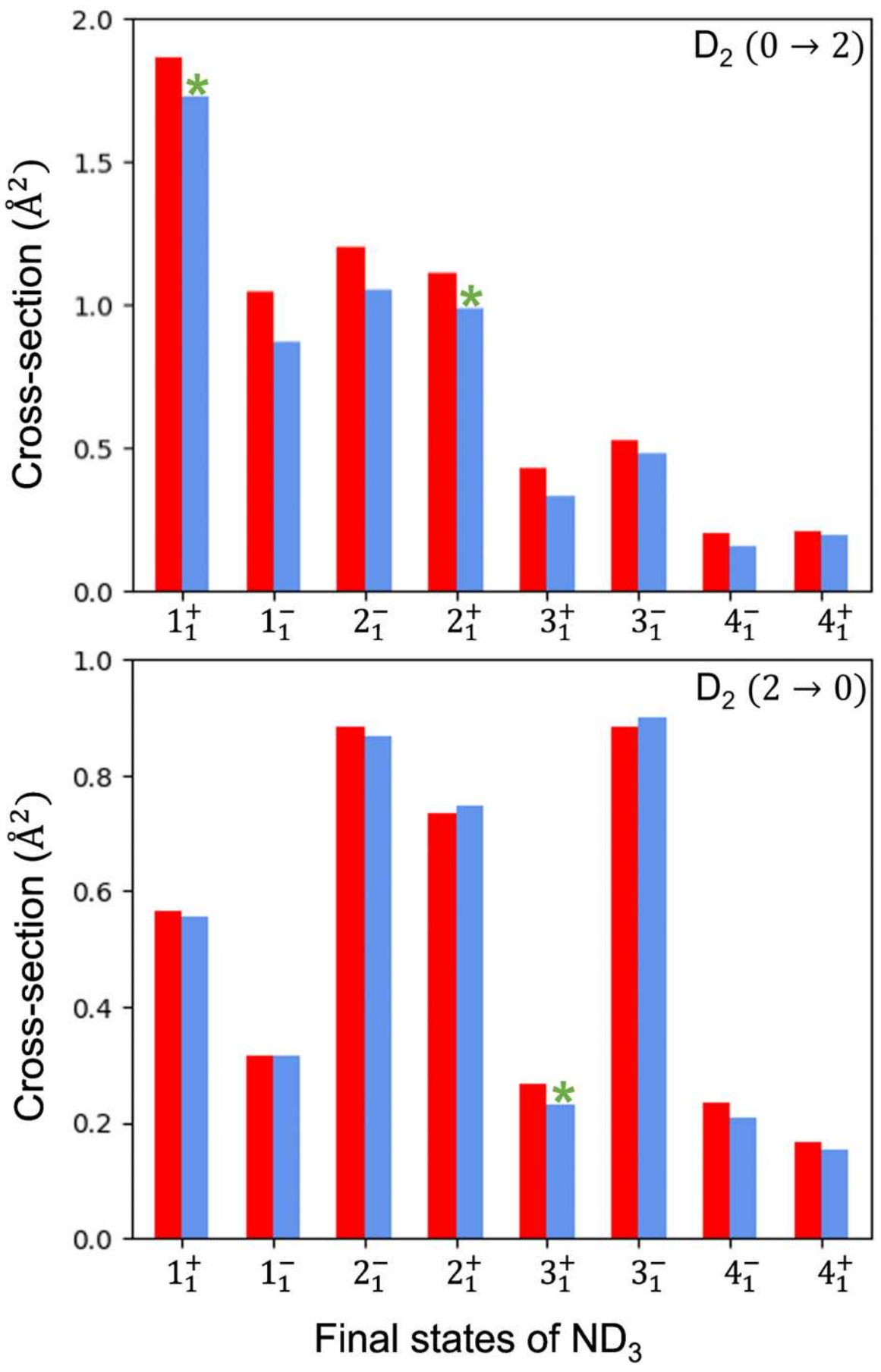

**Figure 6:** Cross sections for excitation of $ND_3(j_k^{\pm})$ states from its ground state $(1_1^-)$ in collisions with $D_2$. In these processes the excitation of $ND_3$ occurs in coincidence with excitation $(0 \to 2$, upper frame) or quenching $(2 \to 0$, lower frame) of the rotational states of $D_2$. Our MQCT results and the full-quantum results of Ref. [40] are shown in blue and red colors, respectively. Transitions labels by asterisk are analyzed in detail in Fig. 2.

Let's take closer look at the process in which for $ND_3$ only the parity changes $(1_1^- \to 1_1^+)$ while $D_2$ is excited or quenched $(0 \leftrightarrow 2)$. The case of $D_2$ excitation exhibits the largest cross section of all (see upper frame of Fig. 6), but the case of $D_2$ quenching exhibits cross section that is a factor of ~ 3 smaller (see lower frame of Fig. 6). The origin of this difference is in the different natures of these two transitions. Namely, the process $(1_1^- 0) \to (1_1^+ 2)$ corresponds to the initial $j_1 = 1$ and $j_2 = 0$, with only one initial value of total angular momentum possible, $j = 1$, that comes with $m = 0$ and $\pm 1$. We checked and found that, for both these components, the transitions are driven by potential coupling with non-zero matrix elements $M_{n'}^{n''}$. In contrast, the process $(1_1^- 2) \to (1_1^+ 0)$ corresponds to the initial $j_1 = 1$ and $j_2 = 2$, with three possible values of total angular momentum, $j = 1, 2$ and 3 that come with their corresponding $m$-states, giving nine different initial $(j, m)$-states. We found that out of these nine components, only four transitions are driven by the potential coupling, while five transitions are driven by Coriolis coupling alone, since the potential coupling is null. The potential-driven components are: $(j, |m|) = (1,1), (2,0), (2,1)$ and $(3,1)$, while the Coriolis-driven components are $(j, |m|) = (1,0), (2,2), (3,0), (3,2)$ and $(3,3)$. On average, this makes the $(1_1^-2) \rightarrow (1_1^+0)$ process much weaker, compared to the $(1_1^-0) \rightarrow (1_1^+2)$ process where each component is driven by the potential coupling (see Fig. 6).

Another interesting feature one may notice in Fig. 6 is that in the upper frame (where $D_2$ is excited) the excitations of $3_1^+$ and $3_1^-$ states exhibit comparable cross sections, while in the lower frame (where $D_2$ is quenched) the excitation of $3_1^-$ state exhibits cross section larger by a factor of three than that for the $3_1^+$ state. We analyzed these processes in detail and found that, first of all, some of potential driven transition probabilities are larger for $(1_1^-2) \rightarrow (3_1^-0)$ transition than for $(1_1^-2) \rightarrow (3_1^+0)$. Moreover, we found that for the $(1_1^-2) \rightarrow (3_1^-0)$ transition, eight out of nine $(j, m)$-components are potential-driven, and only one is Coriolis driven (see Fig. 2), whereas for the $(1_1^-2) \rightarrow (3_1^+0)$ transition seven $(j, m)$-components are potential-driven, while two are Coriolis driven, which makes the later transition weaker (see lower frame of Fig. 6). In contrast, the corresponding transitions in the upper frame of Fig. 6, $(1_1^-0) \rightarrow (3_1^+2)$ and $(1_1^-0) \rightarrow (3_1^-2)$, are characterized by two potential-driven $(j, m)$-components each, and thus exhibit very similar cross sections.

## IV. CONCLUSIONS

In this work we presented the extension of MQCT methodology onto a symmetric-top-rotor + linear-rotor system and applied this theory to describe energy transfer in $ND_3$ + $D_2$ collisions. Calculation of matrix elements for state-to-state transitions driven by potential coupling was carried out in two complementary ways: by direct numerical multi-dimensional integration and by analytic expansion of the PES over a basis set. Two methods were found to give the same results, and a speed-up by a factor of more than an order of magnitude was observed (for the basis set expansion), which offers a fast and reliable method of matrix generation. The derivation of equations is presented in SI, and both methods are available through MQCT suite of codes [[42,43]].

Calculations of energy dependencies of inelastic cross sections were carried out for a large set of state-to-state transitions in $ND_3$ + $D_2$, as presented in SI. It was concluded that the principle of microscopic reversibility is satisfied by MQCT calculations in a broad range of energies, except maybe at the lowest energies, which is not so surprising for an approximate quantum/classical theory. But even in those cases when the deviations from reversibility are observed, they remain manageable (small). We did not find any transition for which the results of MQCT would violate reversibility badly. This concerns transitions of all kinds: when both $ND_3$ and $D_2$ molecules are excited or both are quenched, when one is excited while the other is quenched and vice versa, when $ND_3$ state changes its parity (i.e., a nearly elastic transition between rotationally degenerate states) while $D_2$ is excited or quenched, and when $ND_3$ is excited or quenched while $D_2$ remains in the same state, ground or excited. In all these processes the results of MQCT were found to satisfy the principle of microscopic reversibility.

The analysis of opacity functions (the dependence of inelastic transition probability on collision impact parameter) showed that state-to-state transition processes actively utilize two complementary pathways: potential-driven transitions described by state-to-state transition matrix (within the same $m$-components) and Coriolis-driven transitions (between different $m$-components) that occur in the body-fixed reference frame that rotates during the process of molecule-molecule collision. Transition probabilities that correspond to two pathways exhibit different behavior (as a function of collision impact parameter) with Coriolis-driven processes being somewhat weaker and vanishing completely for head on collisions (zero impact parameter).

A very instructive time-dependent insight was obtained by monitoring the evolution of state populations along MQCT trajectories, for several representative initial conditions. In particular, it was shown that when the initial state of $D_2$ is in its ground state $j = 0$, the excitation of $ND_3$ rotational states proceeds indirectly through the excited $j = 2$ state of $D_2$. In this two-step process the kinetic energy of molecule-molecule collision is first used to excite $D_2(j = 2)$ and only then is transferred to the excited rotational states of $ND_3$. This scenario is presented in the TOC image.

The comparison of our MQCT results with full-quantum results from literature was presented. It was demonstrated that the values of MQCT cross sections follow all trends seen in the full-quantum calculations and are just slightly smaller, by 8% on average. This very good performance of MQCT is encouraging. It opens new opportunities for theoretical description of molecule-molecule collisional energy exchange, since MQCT calculations are quite affordable and offer a very useful time-dependent insight.

## CONFLICTS OF INTEREST

There are no conflicts to declare.

## AKNOWLEDGEMENTS

This research was supported by NSF Chemistry program, grant number CHE-2102465. We used resources of the National Energy Research Scientific Computing Center, which is supported by the Office of Science of the U.S. Department of Energy under Contract No. DE-AC02-5CH11231. Also, we used HPC resources at Marquette funded in part by the National Science foundation award CNS-1828649. Dr. Ad Van der Avoird and Dr. Gerrit C. Groenenboom are acknowledged for stimulating discussions and their continuous support of this project. C. Joy acknowledges the support of Schmitt Fellowship at Marquette. B. Mandal acknowledges the support of Denis J. O'Brien and Eisch Fellowships at Marquette.

*Supplemental Information for:*

# "Mixed Quantum/Classical Theory for Rotational Energy Exchange in Symmetric-Top-Rotor + Linear-Rotor Collisions and a Case Study of $ND_3 + D_2$ System"

by Carolin Joy, Bikramaditya Mandal, Dulat Bostan, and Dmitri Babikov*

*Chemistry Department, Marquette University, Milwaukee, Wisconsin 53201-1881, USA*

## 1. The derivation of expression for transition matrix elements, Eq. (14) in the main text

The wavefunction that describes the motion of the symmetric top + diatom can be expressed analytically as:

$$\Psi_{nm}(\Lambda_1,\Lambda_2) = \sqrt{\frac{2j_1+1}{8\pi^2}}\sqrt{\frac{1}{2(1+\delta_{k_1,0})}} \times \sum_{m_1=-j_1}^{+j_1} \mathrm{C}_{j_1,m_1,j_2,m-m_1}^{j,m}\left[\mathrm{D}_{k_1,m_1}^{j_1*}(\Lambda_1) + \epsilon \mathrm{D}_{-k_1,m_1}^{j_1*}(\Lambda_1)\right]\mathrm{Y}_{m-m_1}^{j_2}(\Lambda_2) \quad \text{(S1)}$$

The multidimensional potential can be expanded over a set of suitable angular functions $\tau_{\lambda_1\mu_1\lambda_2\lambda}(\Lambda_1,\Lambda_2)$ with expansion coefficients $v_{\lambda_1\mu_1\lambda_2\lambda}(R)$ computed by numerical quadrature.

$$V(R,\Lambda_1,\Lambda_2) = \sum_{\lambda_1\mu_1\lambda_2\lambda} v_{\lambda_1\mu_1\lambda_2\lambda}(R) \times \tau_{\lambda_1\mu_1\lambda_2\lambda}(\Lambda_1,\Lambda_2) \quad \text{(S2)}$$

where,

$$\tau_{\lambda_1\mu_1\lambda_2\lambda}(\Lambda_1,\Lambda_2) = \sqrt{\frac{2\lambda_1+1}{4\pi}} \sum_{\eta=-\min(\lambda_1,\lambda_2)}^{+\min(\lambda_1,\lambda_2)} \mathrm{C}_{\lambda_1,\eta,\lambda_2,-\eta}^{\lambda,0}\left[\mathrm{D}_{\mu_1,\eta}^{\lambda_1*}(\Lambda_1) + (-1)^{\lambda_1+\mu_1+\lambda_2+\lambda}\mathrm{D}_{-\mu_1,\eta}^{\lambda_1*}(\Lambda_1)\right]\mathrm{Y}_{-\eta}^{\lambda_2}(\Lambda_2) \quad \text{(S3)}$$

The potential coupling matrix $\mathrm{M}_{n'}^{n''}(R)$ can be expressed as follows:

$$\mathrm{M}_{n'}^{n''}(R) = \langle \Psi_{n''m}(\Lambda_1,\Lambda_2)|V(R,\Lambda_1\Lambda_2)|\Psi_{n'm}(\Lambda_1,\Lambda_2)\rangle \quad \text{(S4)}$$

Here $\Psi_{n''m}(\Lambda_1,\Lambda_2)$ and $\Psi_{n'm}(\Lambda_1,\Lambda_2)$ represents wavefunctions of final and initial states of the molecule.

* Author to whom all correspondence should be addressed; electronic mail: dmitri.babikov@mu.edu

To compute state-to-state matrix elements we have to substitute Eq. (S2) into Eq. (S3), followed by the substitution of Eqs. (S1-S2) on to as Eq. (S4) :

$$\langle \Psi_{n''m}(\Lambda_1,\Lambda_2)|V(R,\Lambda_1\Lambda_2)|\Psi_{n'm}(\Lambda_1,\Lambda_2)\rangle =$$

$$= \sqrt{\frac{2j_1''+1}{8\pi^2}}\sqrt{\frac{1}{2(1+\delta_{k_1'',0})}}\sqrt{\frac{2j_1'+1}{8\pi^2}}\sqrt{\frac{1}{2(1+\delta_{k_1',0})}}\sum_{m_1'=-j_1'}^{+j_1'} \mathrm{C}_{j_1',m_1',j_2',m-m_1'}^{j',m}\sum_{m_1''=-j_1''}^{+j_1''} \mathrm{C}_{j_1'',m_1'',j_2'',m-m_1''}^{j'',m}$$

$$\times \sum_{\lambda_1\mu_1\lambda_2\lambda} \upsilon_{\lambda_1\mu_1\lambda_2\lambda}(R)\sqrt{\frac{2\lambda_1+1}{4\pi}}\sum_{\eta=-\min(\lambda_1,\lambda_2)}^{+\mathrm{mi}\;(\lambda_1,\lambda_2)} \mathrm{C}_{\lambda_1,\eta,\lambda_2,-\eta}^{\lambda,0}\left\langle \mathrm{Y}_{m-m_1''}^{j_2''}(\Lambda_2)\middle|\mathrm{Y}_{-\eta}^{\lambda_2}(\Lambda_2)\middle|\mathrm{Y}_{m-m_1'}^{j_2'}(\Lambda_2)\right\rangle$$

$$\times\begin{bmatrix} \left\langle \mathrm{D}_{k_1'',m_1''}^{j_1''*}(\Lambda_1)\middle|\mathrm{D}_{\mu_1,\eta}^{\lambda_1*}(\Lambda_1)\middle|\mathrm{D}_{k_1',m_1'}^{j_1'*}(\Lambda_1)\right\rangle + (-1)^{\lambda_1+\mu_1+\lambda_2+\lambda}\left\langle \mathrm{D}_{k_1'',m_1''}^{j_1''*}(\Lambda_1)\middle|\mathrm{D}_{-\mu_1,\eta}^{\lambda_1*}(\Lambda_1)\middle|\mathrm{D}_{k_1',m_1'}^{j_1'*}(\Lambda_1)\right\rangle \\ +\epsilon'\left\langle \mathrm{D}_{k_1'',m_1''}^{j_1''*}(\Lambda_1)\middle|\mathrm{D}_{\mu_1,\eta}^{\lambda_1*}(\Lambda_1)\middle|\mathrm{D}_{-k_1',m_1'}^{j_1'*}(\Lambda_1)\right\rangle + (-1)^{\lambda_1+\mu_1+\lambda_2+\lambda}\left\langle \mathrm{D}_{k_1'',m_1''}^{j_1''*}(\Lambda_1)\middle|\mathrm{D}_{-\mu_1,\eta}^{\lambda_1*}(\Lambda_1)\middle|\mathrm{D}_{-k_1',m_1'}^{j_1'*}(\Lambda_1)\right\rangle \\ +\epsilon''\left\langle \mathrm{D}_{-k_1'',m_1''}^{j_1''*}(\Lambda_1)\middle|\mathrm{D}_{\mu_1,\eta}^{\lambda_1*}(\Lambda_1)\middle|\mathrm{D}_{k_1',m_1'}^{j_1'*}(\Lambda_1)\right\rangle + (-1)^{\lambda_1+\mu_1+\lambda_2+\lambda}\left\langle \mathrm{D}_{-k_1'',m_1''}^{j_1''*}(\Lambda_1)\middle|\mathrm{D}_{-\mu_1,\eta}^{\lambda_1*}(\Lambda_1)\middle|\mathrm{D}_{k_1',m_1'}^{j_1'*}(\Lambda_1)\right\rangle \\ +\epsilon'\epsilon''\left\langle \mathrm{D}_{-k_1'',m_1''}^{j_1''*}(\Lambda_1)\middle|\mathrm{D}_{\mu_1,\eta}^{\lambda_1*}(\Lambda_1)\middle|\mathrm{D}_{-k_1',m_1'}^{j_1'*}(\Lambda_1)\right\rangle + (-1)^{\lambda_1+\mu_1+\lambda_2+\lambda}\left\langle \mathrm{D}_{-k_1'',m_1''}^{j_1''*}(\Lambda_1)\middle|\mathrm{D}_{-\mu_1,\eta}^{\lambda_1*}(\Lambda_1)\middle|\mathrm{D}_{-k_1',m_1'}^{j_1'*}(\Lambda_1)\right\rangle \end{bmatrix}$$

For the integrals of spherical harmonics, the following equality can be employed [1]:

$$\langle \mathrm{Y}_{m_3}^{l_3}|\mathrm{Y}_{m_2}^{l_2}\;|\mathrm{Y}_{m_1}^{l_1}\rangle = \sqrt{\frac{(2l_1+1)(2l_2+1)}{4\pi(2l_3+1)}}\,\mathrm{C}_{l_1,0\;,l_2 0}^{l_3,0}\mathrm{C}_{l_1,m_1,l_2,m_2}^{l_3,m_3} \qquad \text{(S5)}$$

For the integrals of Wigner D-functions, the following equality can be employed [1]:

$$\left\langle \mathrm{D}_{m_3m_3'}^{l_3*}\middle|\mathrm{D}_{m_2m_2'}^{l_2*}\middle|\mathrm{D}_{m_1m_1'}^{l_1*}\right\rangle = \frac{8\pi^2}{2l_3+1}\;\mathrm{C}_{l_1,m_1,l_2,m_2}^{l_3,m_3}\mathrm{C}_{l_1,m_1',l_2,m_2'}^{l_3,m_3'} \qquad \text{(S6)}$$

With these, we have:

$$\mathrm{M}_{n'}^{n''}(R) = \sqrt{\frac{2j_1''+1}{8\pi^2}}\sqrt{\frac{1}{2(1+\delta_{k_1'',0})}}\sqrt{\frac{2j_1'+1}{8\pi^2}}\sqrt{\frac{1}{2(1+\delta_{k_1',0})}}\sum_{m_1'=-j_1'}^{+j_1'} \mathrm{C}_{j_1',m_1',j_2',m-m_1'}^{j',m}\sum_{m_1''=-j_1''}^{+j_1''} \mathrm{C}_{j_1'',m_1'',\,j_2'',m-m_1''}^{j'',m}$$

$$\times \sum_{\lambda_1\mu_1\lambda_2\lambda} \upsilon_{\lambda_1\mu_1\lambda_2\lambda}(R) \sqrt{\frac{2\lambda_1+1}{4\pi}} \sum_{\eta=-\mathrm{mi}\ (\lambda_1,\lambda_2)}^{+\mathrm{mi}\ (\lambda_1,\lambda_2)} \mathrm{C}^{\lambda,0}_{\lambda_1,\eta,\,\lambda_2,-\eta} \sqrt{\frac{(2j_2'+1)(2\lambda_2+1)}{4\pi(2j_2''+1)}} \mathrm{C}^{j_2'',0}_{j_2',0,\lambda_2 0} \mathrm{C}^{j_2'',m-m_1''}_{j_2',m-m_1',\lambda_2-\eta}$$

$$\times \frac{8\pi^2}{2j_1''+1} \mathrm{C}^{j_1'',m_1''}_{j_1',m_1',\lambda_1,\eta} \begin{bmatrix} \left(C^{j_1'',k_1''}_{j_1',k_1',\lambda_1,\mu_1} + (-1)^{\lambda_1+\mu_1+\lambda_2+\lambda} C^{j_1'',k_1''}_{j_1',k_1',\lambda_1,-\mu_1}\right) \\ + \epsilon' \left(C^{j_1'',k_1''}_{j_1',-k_1',\lambda_1,\mu_1} + (-1)^{\lambda_1+\mu_1+\lambda_2+\lambda} C^{j_1'',k_1''}_{j_1',-k_1',\lambda_1,-\mu_1}\right) \\ + \epsilon'' \left(C^{j_1'',-k_1''}_{j_1',k_1',\lambda_1,\mu_1} + (-1)^{\lambda_1+\mu_1+\lambda_2+\lambda} C^{j_1'',-k_1''}_{j_1',k_1',\lambda_1,-\mu_1}\right) \\ +\epsilon'\epsilon'' \left(C^{j_1'',-k_1''}_{j_1',-k_1',\lambda_1,\mu_1} + (-1)^{\lambda_1+\mu_1+\lambda_2+\lambda} C^{j_1'',-k_1''}_{j_1',-k_1',\lambda_1,-\mu_1}\right) \end{bmatrix}$$

On further simplification we obtain,

$$\mathrm{M}^{n''}_{n'} = \sqrt{\frac{2j_1'+1}{2j_1''+1}} \sqrt{\frac{2j_2'+1}{2j_2''+1}} \sqrt{\frac{1}{2(1+\delta_{k_1',0})2(1+\delta_{k_1'',0})}} \sum_{\lambda_1\mu_1\lambda_2\lambda} \upsilon_{\lambda_1\mu_1\lambda_2\lambda}(R) \sqrt{\frac{2\lambda_1+1}{4\pi}} \sqrt{\frac{2\lambda_2+1}{4\pi}}$$

$$\times \mathrm{C}^{j_2'',0}_{j_2',0,\lambda_2 0} \sum_{m_1'=-j_1'}^{+j_1'} \mathrm{C}^{j',m}_{j_1',m_1',j_2',m-m_1'} \sum_{m_1''=-j_1''}^{+j_1''} \mathrm{C}^{j'',m}_{j_1'',m_1'',\,j_2'',m-m_1''} \sum_{\eta=-\min(\lambda_1,\lambda_2)}^{+\min(\lambda_1,\lambda_2)} \mathrm{C}^{\lambda,0}_{\lambda_1,\eta,\,\lambda_2,-\eta} \mathrm{C}^{j_2'',m-m_1''}_{j_2',m-m_1',\lambda_2-\eta} \mathrm{C}^{j_1'',m_1''}_{j_1',m_1',\lambda_1,\eta}$$

$$\times \begin{bmatrix} \left(C^{j_1'',k_1''}_{j_1',k_1',\lambda_1,\mu_1} + (-1)^{\lambda_1+\mu_1+\lambda_2+\lambda} C^{j_1'',k_1''}_{j_1',k_1',\lambda_1,-\mu_1}\right) \\ + \epsilon' \left(C^{j_1'',k_1''}_{j_1',-k_1',\lambda_1,\mu_1} + (-1)^{\lambda_1+\mu_1+\lambda_2+\lambda} C^{j_1'',k_1''}_{j_1',-k_1',\lambda_1,-\mu_1}\right) \\ + \epsilon'' \left(C^{j_1'',-k_1''}_{j_1',k_1',\lambda_1,\mu_1} + (-1)^{\lambda_1+\mu_1+\lambda_2+\lambda} C^{j_1'',-k_1''}_{j_1',k_1',\lambda_1,-\mu_1}\right) \\ +\epsilon'\epsilon'' \left(C^{j_1'',-k_1''}_{j_1',-k_1',\lambda_1,\mu_1} + (-1)^{\lambda_1+\mu_1+\lambda_2+\lambda} C^{j_1'',-k_1''}_{j_1',-k_1',\lambda_1,-\mu_1}\right) \end{bmatrix}$$

In general, Clebsch-Gordan coefficients are non-zero only if $m = m_1 + m_2$ and $|j_1 - j_2| \le j \le j_1 + j_2$. Incorporating these properties of CG coefficients [1] we obtain the final state-to-state transition matrix element as follows:

$$\mathrm{M}^{n''}_{n'} = \sqrt{\frac{2j_1'+1}{2j_1''+1}} \sqrt{\frac{2j_2'+1}{2j_2''+1}} \sqrt{\frac{1}{2(1+\delta_{k_1',0})2(1+\delta_{k_1'',0})}} \sum_{\lambda_1\mu_1\lambda_2\lambda} \upsilon_{\lambda_1\mu_1\lambda_2\lambda}(R) \sqrt{\frac{2\lambda_1+1}{4\pi}}$$

$$\times \sqrt{\frac{2\lambda_2+1}{4\pi}} \mathrm{C}^{j_2'',0}_{j_2',0,\lambda_2 0} \sum_{m_1'=-j_1'}^{+j_1'} \mathrm{C}^{j',m}_{j_1',m_1',j_2',m-m_1'} \sum_{\eta=-\mathrm{mi}\ (\lambda_1,\lambda_2)}^{+\min(\lambda_1,\lambda_2)} \mathrm{C}^{j'',m}_{j_1'',m_1'-\eta,\,j_2'',m-(m_1'-\eta)} \mathrm{C}^{\lambda,0}_{\lambda_1,\eta,\,\lambda_2,-\eta} \mathrm{C}^{j_1'',m_1'}_{j_1',m_1'-\eta,\lambda_1,\eta}$$

$$\times \mathrm{C}^{j_2'',m-m_1'}_{j_2',m-(m_1'-\eta),\lambda_2,-\eta} \begin{bmatrix} \left(\mathrm{C}^{j_1'',k_1''}_{j_1',k_1',\lambda_1,\mu_1} + (-1)^{\lambda_1+\mu_1+\lambda_2+\lambda} \mathrm{C}^{j_1'',k_1''}_{j_1',k_1',\lambda_1,-\mu_1}\right) \\ + \epsilon' \left(\mathrm{C}^{j_1'',k_1''}_{j_1',-k_1',\lambda_1,\mu_1} + (-1)^{\lambda_1+\mu_1+\lambda_2+\lambda} \mathrm{C}^{j_1'',k_1''}_{j_1',-k_1',\lambda_1,-\mu_1}\right) \\ + \epsilon'' \left(\mathrm{C}^{j_1'',-k_1''}_{j_1',k_1',\lambda_1,\mu_1} + (-1)^{\lambda_1+\mu_1+\lambda_2+\lambda} \mathrm{C}^{j_1'',-k_1''}_{j_1',k_1',\lambda_1,-\mu_1}\right) \\ +\epsilon'\epsilon'' \left(\mathrm{C}^{j_1'',-k_1''}_{j_1',-k_1',\lambda_1,\mu_1} + (-1)^{\lambda_1+\mu_1+\lambda_2+\lambda} \mathrm{C}^{j_1'',-k_1''}_{j_1',-k_1',\lambda_1,-\mu_1}\right) \end{bmatrix}$$

## 2. Additional figures for publication

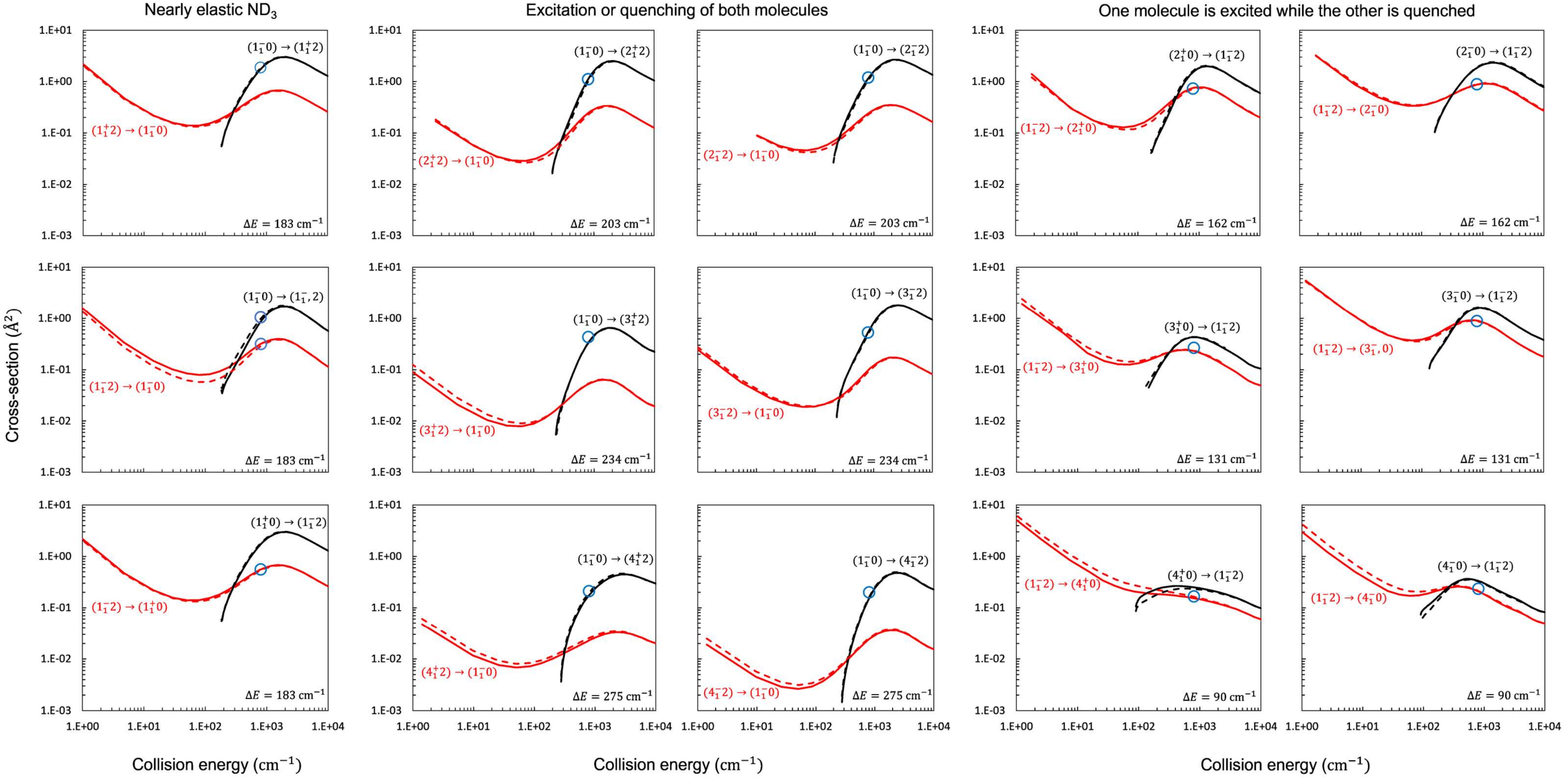

**Figure S1:** The test of microscopic reversibility for transitions between several rotational states of $ND_3 + D_2$ system, labeled as $(j_{1}{}_{k}^{\pm}j_2)$. Cross sections are plotted as a function of collision energy. The data obtained by “direct” MQCT calculations are shown by solid lines, whereas dashed lines represent the results of "reverse" calculations. Red color is used for quenching processes, black color is for excitation processes. Blue symbol indicates full-quantum results of Ref. [2].

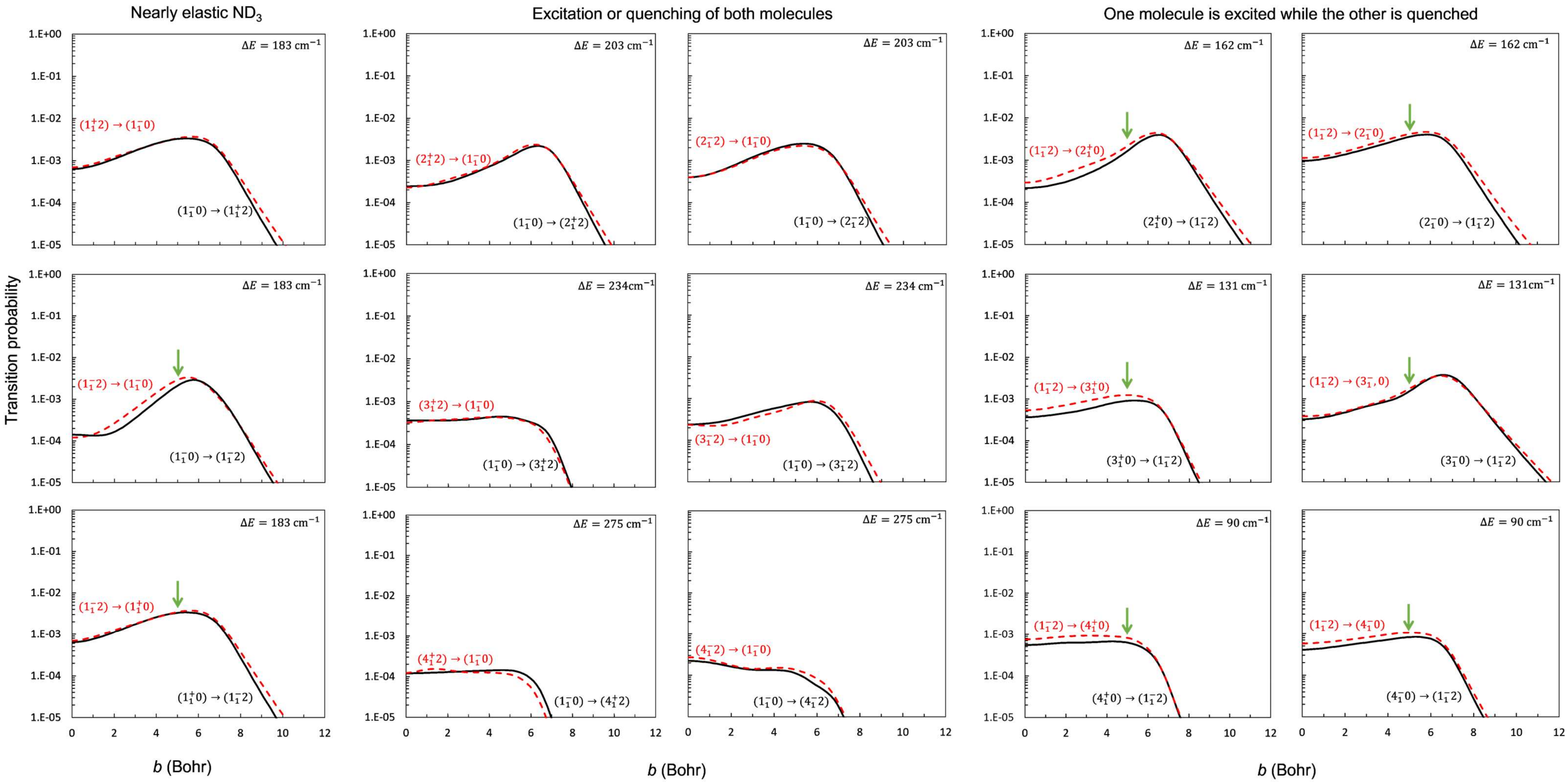

**Figure S2:** Opacity functions for several transitions between the rotational states of $ND_3$ + $D_2$ system, labeled as $(j_{1}{}_{k}^{\pm}j_2)$. Transition probabilities are plotted as a function of collision impact parameter. Collision energy is 800 cm$^{-1}$. Red color is used for quenching processes, black color is for excitation processes. The value of energy difference is given for each transition. Green arrows indicate the value of impact parameter chosen for time-dependent analysis in Fig. 4 of the main text.

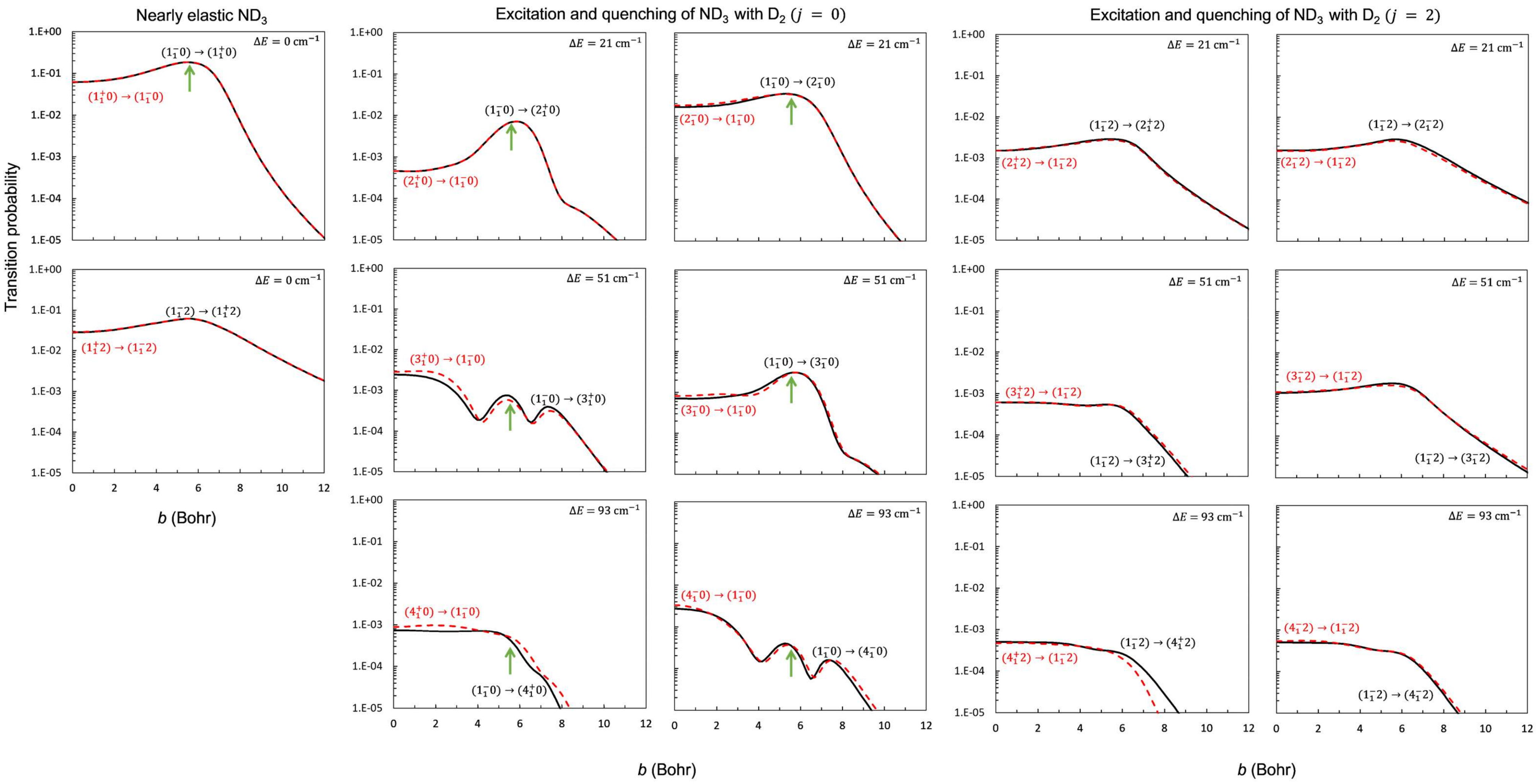

**Figure S3:** Opacity functions for several transitions between the rotational states of $ND_3$ + $D_2$ system, labeled as $(j_{1}{}^{\pm}_{k} j_2)$ with $D_2$ $(j_2)$ remaining in its initial state ($j_2 = 0$ or $j_2 = 2$). Transition probabilities are plotted as a function of collision impact parameter. Collision energy is 800 $cm^{-1}$. Red color is used for quenching processes, black color is for excitation processes. The value of energy difference is given for each transition. Green arrows indicate the value of impact parameter chosen for time-dependent analysis in Fig. 3 of the main text.

**Table S1:** Assignments of total $(j, m)$-components for all transitions discussed in this work in terms of potential coupling and Coriolis coupling. PD is for potential-driven transitions indicated by ✓ mark. If potential coupling is zero (× mark) the transition is driven only by Coriolis coupling. The fraction of Coriolis-driven components is indicated for each transition by red numbers.

| Transition | j | m | PD |
|---|---|---|---|
| $(1_1^+0) \rightarrow (1_1^-2)$ 0/2 | 1 | 0 | ✓ |
| | 1 | 1 | ✓ |
| $\rightarrow (1_1^-0)$ 1/2 | 1 | 0 | × |
| | 1 | 1 | ✓ |
| $(1_1^-0) \rightarrow (1_1^+0)$ 1/2 | 1 | 0 | × |
| | 1 | 1 | ✓ |
| $\rightarrow (2_1^+0)$ 1/2 | 1 | 0 | × |
| | 1 | 1 | ✓ |
| $\rightarrow (2_1^-0)$ 0/2 | 1 | 0 | ✓ |
| | 1 | 1 | ✓ |
| $\rightarrow (3_1^+0)$ 1/2 | 1 | 0 | × |
| | 1 | 1 | ✓ |
| $\rightarrow (3_1^-0)$ 0/2 | 1 | 0 | ✓ |
| | 1 | 1 | ✓ |
| $\rightarrow (4_1^+0)$ 1/2 | 1 | 0 | × |
| | 1 | 1 | ✓ |
| $\rightarrow (4_1^-0)$ 0/2 | 1 | 0 | ✓ |
| | 1 | 1 | ✓ |
| $\rightarrow (1_1^+2)$ 0/2 | 1 | 0 | ✓ |
| | 1 | 1 | ✓ |
| $\rightarrow (1_1^-2)$ 0/2 | 1 | 0 | ✓ |
| | 1 | 1 | ✓ |
| $\rightarrow (2_1^+2)$ 0/2 | 1 | 0 | ✓ |
| | 1 | 1 | ✓ |
| $\rightarrow (2_1^-2)$ 0/2 | 1 | 0 | ✓ |
| | 1 | 1 | ✓ |
| $\rightarrow (3_1^+2)$ 0/2 | 1 | 0 | ✓ |
| | 1 | 1 | ✓ |
| $\rightarrow (3_1^-2)$ 0/2 | 1 | 0 | ✓ |
| | 1 | 1 | ✓ |
| $\rightarrow (4_1^+2)$ 0/2 | 1 | 0 | ✓ |
| | 1 | 1 | ✓ |
| $\rightarrow (4_1^-2)$ 0/2 | 1 | 0 | ✓ |
| | 1 | 1 | ✓ |
| $(1_1^+2) \rightarrow (1_1^-0)$ 5/9 | 1 | 0 | × |
| | 1 | 1 | ✓ |
| | 2 | 0 | ✓ |
| | 2 | 1 | ✓ |
| | 2 | 2 | × |
| | 3 | 0 | × |
| | 3 | 1 | ✓ |
| | 3 | 2 | × |
| | 3 | 3 | × |
| $\rightarrow (1_1^-2)$ 0/9 | 1 | 1 | ✓ |
| | 2 | 0 | ✓ |
| | 2 | 1 | ✓ |
| | 2 | 2 | ✓ |
| | 3 | 0 | ✓ |
| | 3 | 1 | ✓ |
| | 3 | 2 | ✓ |
| | 3 | 3 | ✓ |

| Transition | j | m | PD |
|---|---|---|---|
| $(1_1^-2) \rightarrow (1_1^+0)$ 5/9 | 1 | 0 | × |
| | 1 | 1 | ✓ |
| | 2 | 0 | ✓ |
| | 2 | 1 | ✓ |
| | 2 | 2 | × |
| | 3 | 0 | × |
| | 3 | 1 | ✓ |
| | 3 | 2 | × |
| | 3 | 3 | × |
| $\rightarrow (1_1^-0)$ 4/9 | 1 | 0 | ✓ |
| | 1 | 1 | ✓ |
| | 2 | 0 | × |
| | 2 | 1 | ✓ |
| | 2 | 2 | × |
| | 3 | 0 | ✓ |
| | 3 | 1 | ✓ |
| | 3 | 2 | × |
| | 3 | 3 | × |
| $\rightarrow (2_1^+0)$ 3/9 | 1 | 0 | × |
| | 1 | 1 | ✓ |
| | 2 | 0 | ✓ |
| | 2 | 1 | ✓ |
| | 2 | 2 | ✓ |
| | 3 | 0 | × |
| | 3 | 1 | ✓ |
| | 3 | 2 | ✓ |
| | 3 | 3 | × |
| $\rightarrow (2_1^-0)$ 2/9 | 1 | 0 | ✓ |
| | 1 | 1 | ✓ |
| | 2 | 0 | × |
| | 2 | 1 | ✓ |
| | 2 | 2 | ✓ |
| | 3 | 0 | ✓ |
| | 3 | 1 | ✓ |
| | 3 | 2 | ✓ |
| | 3 | 3 | × |
| $\rightarrow (3_1^+0)$ 2/9 | 1 | 0 | × |
| | 1 | 1 | ✓ |
| | 2 | 0 | ✓ |
| | 2 | 1 | ✓ |
| | 2 | 2 | ✓ |
| | 3 | 0 | × |
| | 3 | 1 | ✓ |
| | 3 | 2 | ✓ |
| | 3 | 3 | ✓ |
| $\rightarrow (3_1^-0)$ 1/9 | 1 | 0 | ✓ |
| | 1 | 1 | ✓ |
| | 2 | 0 | × |
| | 2 | 1 | ✓ |
| | 2 | 2 | ✓ |
| | 3 | 0 | ✓ |
| | 3 | 1 | ✓ |
| | 3 | 2 | ✓ |
| | 3 | 3 | ✓ |
| $\rightarrow (4_1^+0)$ 2/9 | 1 | 0 | × |
| | 1 | 1 | ✓ |
| | 2 | 0 | ✓ |
| | 2 | 1 | ✓ |
| | 2 | 2 | ✓ |
| | 3 | 0 | × |
| | 3 | 1 | ✓ |
| | 3 | 2 | ✓ |
| | 3 | 3 | ✓ |
| $\rightarrow (4_1^-0)$ 1/9 | 1 | 0 | ✓ |
| | 1 | 1 | ✓ |
| | 2 | 0 | × |
| | 2 | 1 | ✓ |
| | 2 | 2 | ✓ |
| | 3 | 0 | ✓ |
| | 3 | 1 | ✓ |
| | 3 | 2 | ✓ |
| | 3 | 3 | ✓ |

| Transition | j | m | PD |
|---|---|---|---|
| $(1_1^-2) \rightarrow (1_1^+2)$ 0/9 | 1 | 0 | ✓ |
| | 1 | 1 | ✓ |
| | 2 | 0 | ✓ |
| | 2 | 1 | ✓ |
| | 2 | 2 | ✓ |
| | 3 | 0 | ✓ |
| | 3 | 1 | ✓ |
| | 3 | 2 | ✓ |
| | 3 | 3 | ✓ |
| $\rightarrow (2_1^+2)$ 0/9 | 1 | 0 | ✓ |
| | 1 | 1 | ✓ |
| | 2 | 0 | ✓ |
| | 2 | 1 | ✓ |
| | 2 | 2 | ✓ |
| | 3 | 0 | ✓ |
| | 3 | 1 | ✓ |
| | 3 | 2 | ✓ |
| | 3 | 3 | ✓ |
| $\rightarrow (2_1^-2)$ 0/9 | 1 | 0 | ✓ |
| | 1 | 1 | ✓ |
| | 2 | 0 | ✓ |
| | 2 | 1 | ✓ |
| | 2 | 2 | ✓ |
| | 3 | 0 | ✓ |
| | 3 | 1 | ✓ |
| | 3 | 2 | ✓ |
| | 3 | 3 | ✓ |
| $\rightarrow (3_1^+2)$ 0/9 | 1 | 0 | ✓ |
| | 1 | 1 | ✓ |
| | 2 | 0 | ✓ |
| | 2 | 1 | ✓ |
| | 2 | 2 | ✓ |
| | 3 | 0 | ✓ |
| | 3 | 1 | ✓ |
| | 3 | 2 | ✓ |
| | 3 | 3 | ✓ |
| $\rightarrow (3_1^-2)$ 0/9 | 1 | 0 | ✓ |
| | 1 | 1 | ✓ |
| | 2 | 0 | ✓ |
| | 2 | 1 | ✓ |
| | 2 | 2 | ✓ |
| | 3 | 0 | ✓ |
| | 3 | 1 | ✓ |
| | 3 | 2 | ✓ |
| | 3 | 3 | ✓ |
| $\rightarrow (4_1^+2)$ 0/9 | 1 | 0 | ✓ |
| | 1 | 1 | ✓ |
| | 2 | 0 | ✓ |
| | 2 | 1 | ✓ |
| | 2 | 2 | ✓ |
| | 3 | 0 | ✓ |
| | 3 | 1 | ✓ |
| | 3 | 2 | ✓ |
| | 3 | 3 | ✓ |
| $\rightarrow (4_1^-2)$ 0/9 | 1 | 0 | ✓ |
| | 1 | 1 | ✓ |
| | 2 | 0 | ✓ |
| | 2 | 1 | ✓ |
| | 2 | 2 | ✓ |
| | 3 | 0 | ✓ |
| | 3 | 1 | ✓ |
| | 3 | 2 | ✓ |
| | 3 | 3 | ✓ |

| Transition | j | m | PD |
|---|---|---|---|
| $(2_1^+0) \rightarrow (1_1^-0)$ 2/3 | 2 | 0 | × |
| | 2 | 1 | ✓ |
| | 2 | 2 | × |
| $\rightarrow (1_1^-2)$ 0/3 | 2 | 0 | ✓ |
| | 2 | 1 | ✓ |
| | 2 | 2 | ✓ |
| $(2_1^-0) \rightarrow (1_1^-0)$ 1/3 | 2 | 0 | ✓ |
| | 2 | 1 | ✓ |
| | 2 | 2 | × |
| $\rightarrow (1_1^-2)$ 0/3 | 2 | 0 | ✓ |
| | 2 | 1 | ✓ |
| | 2 | 2 | ✓ |
| $(3_1^+0) \rightarrow (1_1^-0)$ 3/4 | 3 | 0 | × |
| | 3 | 1 | ✓ |
| | 3 | 2 | × |
| | 3 | 3 | × |
| $\rightarrow (1_1^-2)$ 0/4 | 3 | 0 | ✓ |
| | 3 | 1 | ✓ |
| | 3 | 2 | ✓ |
| | 3 | 3 | ✓ |
| $(3_1^-0) \rightarrow (1_1^-0)$ 2/4 | 3 | 0 | ✓ |
| | 3 | 1 | ✓ |
| | 3 | 2 | × |
| | 3 | 3 | × |
| $\rightarrow (1_1^-2)$ 0/4 | 3 | 0 | ✓ |
| | 3 | 1 | ✓ |
| | 3 | 2 | ✓ |
| | 3 | 3 | ✓ |
| $(4_1^+0) \rightarrow (1_1^-0)$ 4/5 | 4 | 0 | × |
| | 4 | 1 | ✓ |
| | 4 | 2 | × |
| | 4 | 3 | × |
| | 4 | 4 | × |
| $\rightarrow (1_1^-2)$ 1/5 | 4 | 0 | ✓ |
| | 4 | 1 | ✓ |
| | 4 | 2 | ✓ |
| | 4 | 3 | ✓ |
| | 4 | 4 | × |
| $(4_1^-0) \rightarrow (1_1^-0)$ 3/5 | 4 | 0 | ✓ |
| | 4 | 1 | ✓ |
| | 4 | 2 | × |
| | 4 | 3 | × |
| | 4 | 4 | × |
| $\rightarrow (1_1^-2)$ 1/5 | 4 | 0 | ✓ |
| | 4 | 1 | ✓ |
| | 4 | 2 | ✓ |
| | 4 | 3 | ✓ |
| | 4 | 4 | × |

**Table S1:** *Continued*

| Transition | j | m | PD |
|---|---|---|---|
| $(2_1^+2) \rightarrow (1_1^-0)$ 9/15 | 0 | 0 | × |
| | 1 | 0 | ✓ |
| | 1 | 1 | ✓ |
| | 2 | 0 | × |
| | 2 | 1 | ✓ |
| | 2 | 2 | × |
| | 3 | 0 | ✓ |
| | 3 | 1 | ✓ |
| | 3 | 2 | × |
| | 3 | 3 | × |
| | 4 | 0 | × |
| | 4 | 1 | ✓ |
| | 4 | 2 | × |
| | 4 | 3 | × |
| | 4 | 4 | × |
| $\rightarrow (1_1^-2)$ 1/15 | 0 | 0 | ✓ |
| | 1 | 0 | ✓ |
| | 1 | 1 | ✓ |
| | 2 | 0 | ✓ |
| | 2 | 1 | ✓ |
| | 2 | 2 | ✓ |
| | 3 | 0 | ✓ |
| | 3 | 1 | ✓ |
| | 3 | 2 | ✓ |
| | 3 | 3 | ✓ |
| | 4 | 0 | ✓ |
| | 4 | 1 | ✓ |
| | 4 | 2 | ✓ |
| | 4 | 3 | ✓ |
| | 4 | 4 | × |
| $(2_1^-2) \rightarrow (1_1^-0)$ 8/15 | 0 | 0 | ✓ |
| | 1 | 0 | × |
| | 1 | 1 | ✓ |
| | 2 | 0 | ✓ |
| | 2 | 1 | ✓ |
| | 2 | 2 | × |
| | 3 | 0 | × |
| | 3 | 1 | ✓ |
| | 3 | 2 | × |
| | 3 | 3 | × |
| | 4 | 0 | ✓ |
| | 4 | 1 | ✓ |
| | 4 | 2 | × |
| | 4 | 3 | × |
| | 4 | 4 | × |
| $\rightarrow (1_1^-2)$ 1/15 | 0 | 0 | ✓ |
| | 1 | 0 | ✓ |
| | 1 | 1 | ✓ |
| | 2 | 0 | ✓ |
| | 2 | 1 | ✓ |
| | 2 | 2 | ✓ |
| | 3 | 0 | ✓ |
| | 3 | 1 | ✓ |
| | 3 | 2 | ✓ |
| | 3 | 3 | ✓ |
| | 4 | 0 | ✓ |
| | 4 | 1 | ✓ |
| | 4 | 2 | ✓ |
| | 4 | 3 | ✓ |
| | 4 | 4 | × |

| Transition | j | m | PD |
|---|---|---|---|
| $(3_1^+2) \rightarrow (1_1^-0)$ 13/20 | 1 | 0 | × |
| | 1 | 1 | ✓ |
| | 2 | 0 | ✓ |
| | 2 | 1 | ✓ |
| | 2 | 2 | × |
| | 3 | 0 | × |
| | 3 | 1 | ✓ |
| | 3 | 2 | × |
| | 3 | 3 | × |
| | 4 | 0 | ✓ |
| | 4 | 1 | ✓ |
| | 4 | 2 | × |
| | 4 | 3 | × |
| | 4 | 4 | × |
| | 5 | 0 | × |
| | 5 | 1 | ✓ |
| | 5 | 2 | × |
| | 5 | 3 | × |
| | 5 | 4 | × |
| | 5 | 5 | × |
| $\rightarrow (1_1^-2)$ 3/20 | 1 | 0 | ✓ |
| | 1 | 1 | ✓ |
| | 2 | 0 | ✓ |
| | 2 | 1 | ✓ |
| | 2 | 2 | ✓ |
| | 3 | 0 | ✓ |
| | 3 | 1 | ✓ |
| | 3 | 2 | ✓ |
| | 3 | 3 | ✓ |
| | 4 | 0 | ✓ |
| | 4 | 1 | ✓ |
| | 4 | 2 | ✓ |
| | 4 | 3 | ✓ |
| | 4 | 4 | × |
| | 5 | 0 | ✓ |
| | 5 | 1 | ✓ |
| | 5 | 2 | ✓ |
| | 5 | 3 | ✓ |
| | 5 | 4 | × |
| | 5 | 5 | × |

| Transition | j | m | PD |
|---|---|---|---|
| $(3_1^-2) \rightarrow (1_1^-0)$ 12/20 | 1 | 0 | ✓ |
| | 1 | 1 | ✓ |
| | 2 | 0 | × |
| | 2 | 1 | ✓ |
| | 2 | 2 | × |
| | 3 | 0 | ✓ |
| | 3 | 1 | ✓ |
| | 3 | 2 | × |
| | 3 | 3 | × |
| | 4 | 0 | × |
| | 4 | 1 | ✓ |
| | 4 | 2 | × |
| | 4 | 3 | × |
| | 4 | 4 | × |
| | 5 | 0 | ✓ |
| | 5 | 1 | ✓ |
| | 5 | 2 | × |
| | 5 | 3 | × |
| | 5 | 4 | × |
| | 5 | 5 | × |
| $\rightarrow (1_1^-2)$ 3/20 | 1 | 0 | ✓ |
| | 1 | 1 | ✓ |
| | 2 | 0 | ✓ |
| | 2 | 1 | ✓ |
| | 2 | 2 | ✓ |
| | 3 | 0 | ✓ |
| | 3 | 1 | ✓ |
| | 3 | 2 | ✓ |
| | 3 | 3 | ✓ |
| | 4 | 0 | ✓ |
| | 4 | 1 | ✓ |
| | 4 | 2 | ✓ |
| | 4 | 3 | ✓ |
| | 4 | 4 | × |
| | 5 | 0 | ✓ |
| | 5 | 1 | ✓ |
| | 5 | 2 | ✓ |
| | 5 | 3 | ✓ |
| | 5 | 4 | × |
| | 5 | 5 | × |

| Transition | j | m | PD |
|---|---|---|---|
| $(4_1^+2) \rightarrow (1_1^-0)$ 18/20 | 2 | 0 | × |
| | 2 | 1 | ✓ |
| | 2 | 2 | × |
| | 3 | 0 | ✓ |
| | 3 | 1 | ✓ |
| | 3 | 2 | × |
| | 3 | 3 | × |
| | 4 | 0 | × |
| | 4 | 1 | ✓ |
| | 4 | 2 | × |
| | 4 | 3 | × |
| | 4 | 4 | × |
| | 5 | 0 | ✓ |
| | 5 | 1 | ✓ |
| | 5 | 2 | × |
| | 5 | 3 | × |
| | 5 | 4 | × |
| | 5 | 5 | × |
| | 6 | 0 | × |
| | 6 | 1 | ✓ |
| | 6 | 2 | × |
| | 6 | 3 | × |
| | 6 | 4 | × |
| | 6 | 5 | × |
| | 6 | 6 | × |
| $\rightarrow (1_1^-2)$ 6/25 | 2 | 0 | ✓ |
| | 2 | 1 | ✓ |
| | 2 | 2 | ✓ |
| | 3 | 0 | ✓ |
| | 3 | 1 | ✓ |
| | 3 | 2 | ✓ |
| | 3 | 3 | ✓ |
| | 4 | 0 | ✓ |
| | 4 | 1 | ✓ |
| | 4 | 2 | ✓ |
| | 4 | 3 | ✓ |
| | 4 | 4 | × |
| | 5 | 0 | ✓ |
| | 5 | 1 | ✓ |
| | 5 | 2 | ✓ |
| | 5 | 3 | ✓ |
| | 5 | 4 | × |
| | 5 | 5 | × |
| | 6 | 0 | ✓ |
| | 6 | 1 | ✓ |
| | 6 | 2 | ✓ |
| | 6 | 3 | ✓ |
| | 6 | 4 | × |
| | 6 | 5 | × |
| | 6 | 6 | × |

| Transition | j | m | PD |
|---|---|---|---|
| $(4_1^-2) \rightarrow (1_1^-0)$ 17/20 | 2 | 0 | ✓ |
| | 2 | 1 | ✓ |
| | 2 | 2 | × |
| | 3 | 0 | × |
| | 3 | 1 | ✓ |
| | 3 | 2 | × |
| | 3 | 3 | × |
| | 4 | 0 | ✓ |
| | 4 | 1 | ✓ |
| | 4 | 2 | × |
| | 4 | 3 | × |
| | 4 | 4 | × |
| | 5 | 0 | × |
| | 5 | 1 | ✓ |
| | 5 | 2 | × |
| | 5 | 3 | × |
| | 5 | 4 | × |
| | 5 | 5 | × |
| | 6 | 0 | ✓ |
| | 6 | 1 | ✓ |
| | 6 | 2 | × |
| | 6 | 3 | × |
| | 6 | 4 | × |
| | 6 | 5 | × |
| | 6 | 6 | × |
| $\rightarrow (1_1^-2)$ 6/25 | 2 | 0 | ✓ |
| | 2 | 1 | ✓ |
| | 2 | 2 | ✓ |
| | 3 | 0 | ✓ |
| | 3 | 1 | ✓ |
| | 3 | 2 | ✓ |
| | 3 | 3 | ✓ |
| | 4 | 0 | ✓ |
| | 4 | 1 | ✓ |
| | 4 | 2 | ✓ |
| | 4 | 3 | ✓ |
| | 4 | 4 | × |
| | 5 | 0 | ✓ |
| | 5 | 1 | ✓ |
| | 5 | 2 | ✓ |
| | 5 | 3 | ✓ |
| | 5 | 4 | × |
| | 5 | 5 | × |
| | 6 | 0 | ✓ |
| | 6 | 1 | ✓ |
| | 6 | 2 | ✓ |
| | 6 | 3 | ✓ |
| | 6 | 4 | × |
| | 6 | 5 | × |
| | 6 | 6 | × |